\RequirePackage{fix-cm}
\documentclass[smallextended]{svjour3}       

\usepackage{amsfonts}
\usepackage{amsmath}
\usepackage{amssymb}
\usepackage{bm}
\usepackage{caption}
\usepackage{cite}
\usepackage{color}
\usepackage{graphicx}
\usepackage[dvips,colorlinks,bookmarksopen,bookmarksnumbered,citecolor=red,urlcolor=red]{hyperref}
\usepackage{moreverb}
\usepackage{multirow}
\usepackage{psfrag}
\usepackage{subcaption}
\usepackage{xcolor}

\graphicspath{ {./images/} }

\def\volumeyear{2013}

\def \grad{\nabla}
\def \p{\partial}

\def \CC{\mathbb{C}}
\def \FF{\mathbb{F}}
\def \II{\mathbb{I}}
\def \RR{\mathbb{R}}
\def \PP{\mathbb{P}}

\def \F{\vec{F}}
\def \G{\vec{G}}
\def \N{\vec{N}}
\def \U{\vec{U}}
\def \V{\vec{V}}
\def \X{\vec{X}}
\def \f{\vec{f}}
\def \g{\vec{g}}
\def \q{\vec{q}}

\def \u{\vec{u}}
\def \vchi{\vec{\chi}}
\def \vphi{\vec{\phi}}

\def \x{\vec{x}}
\def \ps{p_\text{s}}
\def \betas{\beta_\text{s}}

\def \Omegaf{\Omega^\text{f}}
\def \Omegal{\Omega^\text{l}}
\def \Omegaw{\Omega^\text{w}}

\def \Dx{\mbox{d} \x}
\def \DA{\mbox{d} A(\X)}
\def \DX{\mbox{d} \X}
\def \Dq{\mbox{d} \q}

\def \half{\frac{1}{2}}
\def \dq{\Delta \q}
\def \dt{\Delta t}
\def \dx{\Delta \x}

\def \cR{\vec{\mathcal R}}
\def \cRl{\cR^\text{l}}
\def \cRw{\cR^\text{w}}
\def \cS{\vec{\mathcal S}}
\def \cSl{\cS^\text{l}}
\def \cSw{\cS^\text{w}}

\newcommand{\tr}{\mathop{\mathrm{tr}}}

\renewcommand{\d}[2]{\frac{\mathrm{d} #1}{\mathrm{d} #2}}
\newcommand{\D}[2]{\frac{\p #1}{\p #2}}
\newcommand{\DD}[2]{\frac{\p^2 #1}{\p #2^2}}
\renewcommand{\vec}[1]{\bm{\mathrm{#1}}}

\begin{document}

\title{Immersed boundary-finite element model of fluid-structure interaction in the aortic root}
\titlerunning{Immersed boundary-finite element model of FSI in the aortic root}

\author{Vittoria Flamini \and Abe DeAnda \and Boyce E.~Griffith}

\institute{
  V.~Flamini \at
  Department of Mechanical and Aerospace Engineering\\
  New York University Polytechnic School of Engineering\\
  Brooklyn, New York, USA
  \and
  A.~DeAnda \at
  Department of Cardiothoracic Surgery\\
  New York University School of Medicine\\
  New York, New York, USA
  \and
  B.~E.~Griffith \at
  Departments of Mathematics and Biomedical Engineering and McAllister Heart Institute\\
  Phillips Hall, Campus Box 3250\\
  University of North Carolina\\
  Chapel Hill, North Carolina, USA\\
  Phone:~(919)~962-1294\\
  \email{boyceg@email.unc.edu}}

\date{Received: date / Accepted: date}

\maketitle

\begin{abstract}
  It has long been recognized that aortic root elasticity helps to
  ensure efficient aortic valve closure, but our understanding of the
  functional importance of the elasticity and geometry of the aortic
  root continues to evolve as increasingly detailed in vivo imaging
  data become available.  Herein, we describe fluid-structure
  interaction models of the aortic root, including the aortic valve
  leaflets, the sinuses of Valsalva, the aortic annulus, and the
  sinotubular junction, that employ a version of Peskin's immersed
  boundary (IB) method with a finite element (FE) description of the
  structural elasticity.  We develop both an idealized model of the
  root with three-fold symmetry of the aortic sinuses and valve
  leaflets, and a more realistic model that accounts for the
  differences in the sizes of the left, right, and noncoronary sinuses
  and corresponding valve cusps.  As in earlier work, we use
  fiber-based models of the valve leaflets, but this study extends
  earlier IB models of the aortic root by employing incompressible
  hyperelastic models of the mechanics of the sinuses and ascending
  aorta using a constitutive law fit to experimental data from human
  aortic root tissue.  In vivo pressure loading is accounted for by a
  backwards displacement method that determines the unloaded
  configurations of the root models.  Our models yield realistic
  cardiac output at physiological pressures, with low transvalvular
  pressure differences during forward flow, minimal regurgitation
  during valve closure, and realistic pressure loads when the valve is
  closed during diastole.  Further, results from high-resolution
  computations demonstrate that IB models of the aortic valve are able
  to produce essentially grid-converged dynamics at practical grid
  spacings for the high-Reynolds number flows of the aortic root.

\end{abstract}

\section{Introduction}

The aortic root consists of the three semilunar aortic valve cusps,
the three sinuses of Valsalva, which are bulbous cavities positioned
behind each valve leaflet, the aortic annulus, where the aortic valve
leaflets attach to the aorta, and the sinotubular junction, where the
sinuses merge into the ascending aorta.  The healthy aortic valve acts to
ensure unidirectional flow of oxygenated blood from the heart to the
tissues of the body, including to the heart itself via the coronary
circulation.  Diseases of the aortic valve can result in stenosis or
regurgitation, and severe aortic valve disease is treated by repairing
or by replacing the diseased valve with either a mechanical or a
bioprosthetic valve \cite{carr2005}.  Approximately 50,000 such
procedures are performed each year in the United States
\cite{freeman2005, bonow2006, singh2008}.  Continuing advances in
noninvasive in vivo imaging of blood flow and tissue deformation,
mechanical tests specifically targeted to biological tissues, and
computer modeling and simulation enable integrative studies of the
dynamics and mechanics of the aortic root \cite{humphrey2002}.  Such
work has the potential to improve both medical devices and surgical
procedures used to treat patients with valvular heart disease, and
also clinical approaches to patient risk assessment and treatment
planning.

Aortic compliance, and specifically the compliance of the aortic
sinuses, has a fundamental role in the function of the aortic root.
The sinuses act as reservoirs, storing blood during systole and then
releasing it in diastole to facilitate flow in the coronaries
\cite{thubrikar1989}.  The compliance of the sinuses also helps to
ensure the proper closure of the aortic valve \cite{bellhouse1968}.
Further, the shape of the aortic root causes blood to circulate in the
sinuses, generating vortices that act both on the valve leaflets and
on the sinus walls.  These vortices generate forces that facilitate
effective valve closure and act as a regulatory mechanism that
synchronizes closure \cite{bellhouse1968}.  Although this mechanism of
efficient valve closure was first postulated by Leonardo da Vinci,
there remains some debate in the surgical community about whether it
is important to recreate the anatomic geometry of the native aortic
root in aortic valve and root replacement procedures \cite{cheng2007}.

More recently, our understanding of the functional importance of the
elasticity and anatomical geometry of the aortic root has evolved as
increasingly detailed data have been acquired on the in vivo
deformations that occur during the cardiac cycle.  For instance, it
has been shown that the aortic root undergoes a multi-modal series of
conformational changes even before the leaflets open \cite{dagum1999}.
It has been argued that these deformations act to reduce shear
stresses on the valve leaflets and thereby prolong the life of the
native valve leaflets \cite{cheng2007}.  Annular and commissural
flexibility may be a key component in this interaction, and annular
flexibility is lost with the implantation of a stented artificial
valve, possibly reducing the lifetime of bioprosthetic valve leaflets.
The annular commissures also participate in load sharing, which
reduces peak stresses on the aortic cusps immediately after valve
closure.  Devices such as the Medtronic 3f Aortic Bioprosthesis
attempt to account for both the annular and the commissural
flexibility of the native root, but despite the theoretical advantages
offered by such designs, more traditional devices with rigid annuluses
still dominate clinical practice.


A challenge in modeling and simulating the mechanical response of
arteries, including the aorta, is that the arterial walls are
continuously subjected to substantial intraluminal pressures in their
in vivo state.  Thus, arterial geometries that are derived from in
vivo imaging generally correspond to a loaded configuration.  In
addition, as first demonstrated by Fung \cite{fung1983} and by
Vaishnav and Vossoughi \cite{vaishnav1983}, even the unloaded
configuration of the artery is not stress free.  Instead, because of
tissue growth and remodeling, the arterial wall includes residual
stresses and is subject to axial tethering \cite{fung1991,
  humphrey2002, cardamone2009}. These factors all conspire to make
determining the unloaded arterial geometry challenging. Indeed, both
the deformation due to intraluminal pressure loading and the residual
stresses cannot be measured directly in vivo, and therefore these
quantities must be estimated.  Residual stresses may be determined
from ex vivo experiments, whereas the zero-pressure configuration of
blood vessels can be determined by solving the inverse elastostatic
problem.

Different approaches have been developed to solve the inverse
elastostatic problem for complex arterial geometries derived from
medical imaging data.  Most of these strategies can be grouped in two
broad categories.  One group of approaches focuses on retrieving the
initial deformation field and the initial stress field of the artery
while keeping the imaging-derived geometry unaltered.  An example of
this type of approach is the modified updated Lagrangian formulation
(MULF) introduced by Gee et al.~\cite{gee2009}, which is an
incremental prestressing method that recovers the equilibrium
configuration rather than the zero-pressure geometry.  A second group
of approaches determines an unloaded configuration of the vessel that,
when subjected to intraluminal pressure loading, will inflate to match
the imaging-derived geometry.  Among the first to develop a stationary
method to solve the inverse elastostatic problem were Govindjee and
Mihalic \cite{govindjee1998}.  This method was extended by Lu et
al.~\cite{lu2007} to arteries, but this approach does not appear to
have been widely used in practice, possibly because it requires a
modification in the finite element (FE) solution scheme that renders
this method somewhat difficult to implement within commercial FE
analysis software.  In comparison, the backward incremental method
developed by de Putter et al.~\cite{dePutter2007} requires updating
the deformation gradient and thereby can be implemented with fewer
modifications to the finite element code.  Perhaps the most
straightforward approach to retrieving the zero-pressure configuration
is the backward displacement method proposed by Bols et
al.~\cite{bols2012}, which iteratively modifies the coordinates of the
reference configuration rather than the deformation gradient tensor.
We use the backward displacement method in this work to recover the
unloaded configuration of an idealized model of the aortic root.

While the methods to obtain an estimation of the zero-pressure
configuration of a blood vessel can be applied to any geometry,
determining residual stresses is a local, vessel-specific task that
presently requires harvesting and then destroying the blood vessel.
Computational approaches to accounting for residual stresses are
usually based on an assumed `open' configuration, which is
subsequently `closed,' thereby reversing the process that releases
residual stresses \cite{alastrue2007,creane2012}.  This approach is
difficult to apply to noncylindrical geometries and, in particular,
cannot be readily applied to imaging-derived geometries
\cite{creane2012}.
A more generic approach to including residual stress in arteries is
described in Alastru\'{e} et al.~\cite{alastrue2007} and is based on a
Kr\"{o}ner-Lee decomposition of the deformation gradient tensor.  For
simplicity, we do not consider residual stresses in the present work,
although we do account for initial strains resulting from in vivo
pressure loading.

Most the methods surveyed above have been developed and applied to
study aneurysmal arteries \cite{lu2007,gee2009}, although applications
to normal vessels as well as other pathophysiological conditions are
possible.  Because arterial wall stress distributions are considered
good predictors of aneurysm rupture, improvements in the accuracy of
computed wall stress distributions are expected to facilitate
improvements in aneurysm rupture prediction.  In contrast, although
the important role of aortic root compliance in valve closure is well
documented \cite{croft2010,crosetto2011}, many dynamic analyses of the
aortic valve tend to focus on the mechanical behavior of the valve
leaflets and often model the sinuses and the ascending aorta as rigid
\cite{griffith2009,viscardi2010,griffith2012,yao2012} or as linear
elastic materials \cite{conti2010,marom2012}, but experimental data
\cite{azadani2012} show that the sinuses and the ascending aorta
exhibit a nonlinear hyperelastic behavior.  Examples of nonlinear
hyperelastic constitutive models used in fluid-structure interaction
simulations of the aortic root can be found in studies by De Hart et
al.~\cite{dehart2003} and by Weinberg and Mofrad \cite{weinberg2008},
which both adopt an arbitrary Lagrangian-Eulerian (ALE) approach.  The
application of ALE fluid-structure interaction schemes to modeling
aortic valve dynamics is somewhat limited by the remeshing required by
such methods as the structure deforms
\cite{sotiropoulos2009}. Attempts to tackle this limitation have
included restricting the problem to a two-dimensional analysis
\cite{dumont2004} and the implementation of complex remeshing
algorithms in three-dimensional analyses \cite{cheng2004}.

An alternative approach to modeling interactions between the blood and
the aortic valve and root is offered by Peskin's immersed boundary (IB)
method \cite{peskin2002}, which was introduced to simulate
cardiac valve dynamics \cite{peskin1972} and was subsequently extended
to model fluid-structure interaction in
the heart and its valves
\cite{peskin1996,mcqueen1997,mcqueen2000,kovacs2001,mcqueen2001,griffith2007,griffith2009-ib_chapter,griffith2009,griffith2012,luo2012,ma2013}.
This IB approach to fluid-structure interaction is to describe
structural stresses and deformations in Lagrangian form, and to
describe the momentum, viscosity, and incompressibility of the coupled
fluid-structure system in Eulerian form.  Integral transforms with
Dirac delta function kernels mediate interaction between Lagrangian
and Eulerian variables.  When discretized, the IB formulation of the
equations of motion replaces these singular kernel functions with
regularized kernels that are designed to ensure conservation of
physical quantities such as force and torque when converting between
Lagrangian and Eulerian forms.  This numerical approach allows the
Lagrangian structural mesh to overlay the background Eulerian grid in
an arbitrary manner, thereby avoiding the need to deploy dynamic
body-fitted grids.  In addition, because the immersed structures move
according to a common interpolated velocity field, the IB method
offers an implicit contact model that prevents the valve leaflets from
interpenetrating, even when subjected to substantial diastolic
pressure loads \cite{griffith2009, griffith2012}.  Related
sharp-interface IB methods have also been developed and used by
Borazjani, Ge, and Sotiropoulos to simulate valvular dynamics for
models of both rigid and flexible aortic valve prostheses
\cite{borazjani2008,borazjani2013}.  Previous work demonstrated that three-dimensional IB models of the aortic
valve can yield physiological cardiac output at realistic pressures
\cite{griffith2009, griffith2012}.  However, we are aware of no
previous study that demonstrates that the IB method yields reasonably
well-resolved simulation results for flexible aortic valve models in
three spatial dimensions at practical grid spacings.

The primary aim of this study is to develop new fluid-structure interaction models of the
aortic root that substantially extend earlier IB models of aortic
valve dynamics \cite{griffith2009, griffith2012} by including
descriptions of the elasticity of the aortic sinuses and the ascending
aorta.  Herein, the aortic sinuses and proximal ascending aorta are
modeled as an incompressible, isotropic, hyperelastic material with an
exponential neo-Hookean strain-energy functional \cite{humphrey2002,
  delfino1997} that we fit to experimental data obtained by Azadani et
al.~\cite{azadani2012} from human aortic root tissue samples.  A
separate finite element analysis is employed to estimate the unloaded
aortic geometry using a method based on the backward displacement
method of Bols et al.~\cite{bols2012}.  Our fluid-structure
interaction simulations employ hybrid models of the aortic root that
use a fiber-based description of the thin aortic valve leaflets, as
done in previous studies \cite{griffith2009, griffith2012}, along with
a finite element-based description \cite{griffith-submitted} of the
comparatively thick walls of the aortic sinuses and proximal ascending
aorta.

Two related aortic root models are considered in this work.  Each
captures the complex interactions between the flow, the thin flexible
aortic valve cusps, and the deformable walls of the aortic sinuses and
the ascending aorta.  The first model employs a highly idealized
anatomical geometry, in which the initial and reference configurations
of the aortic sinuses and valve cusps are assumed to exhibit
three-fold symmetry.  The second model employs a more realistic
description of the aortic root anatomy that accounts for the
differences in the sizes of the left, right, and noncoronary sinuses.
Both models are fully three dimensional in the sense that there are no
symmetry conditions imposed on the subsequent fluid dynamics or
structural kinematics.  Indeed, because of the relatively high
Reynolds numbers of the systolic flow, symmetry breaking causes the
three leaflets to move asynchronously even in the highly idealized
aortic root geometry.

Using the idealized model, we perform a refinement study to
demonstrate that the present methods are able to yield essentially
grid-converged results at practical grid spacings.  Specifically,
under grid refinement, only relatively small differences are observed
in stroke volume, maximum flow velocity, and vessel distensibility and
stress distributions.  Relatively large differences remain in the
details of the flow in the vicinity of the valve cusps, and in the
details of the kinematics of the valve cusps.  Higher spatial
resolution is likely needed to resolve fully the fine-scale dynamics
of the valve during systole.  Nonetheless, a key contribution of this
study is that it demonstrates, for the first time, that even bulk flow
properties are reasonably resolved by IB models of aortic valve
dynamics at practical spatial resolutions.

We also compare the numerical predictions of the symmetric and
asymmetric aortic root models. Previous studies have considered the
influence of asymmetric leaflets, such as bileaflet aortic valves, in
symmetric aortic root configurations \cite{croft2010,weinberg2008}.
In this work, we focus on assessing how symmetric and asymmetric
geometries affect bulk flow hemodynamics and aortic wall mechanics. In
particular, we aim to determine the extent that a symmetric geometry
can be considered a reliable model of the aortic root, and to
determine situations where an asymmetric geometry gives a more
physiologically accurate results.


In our dynamic analyses, we pace the aortic root model to an
essentially periodic steady state using a prescribed, periodic
left-ventricular pressure waveform.  A Windkessel model provides
downstream loading for the aortic root.  Notice that in these models,
flow rates are not prescribed, but rather emerge from the computation.
In all cases, physiological cardiac output is obtained at
physiological driving and loading pressures, with low transvalvular
pressure differences during forward flow, minimal regurgitation during
valve closure, and realistic transvalvular pressure loads when the
valve is closed during diastole.

\section{Methods}

\subsection{Immersed Boundary Formulation}

The immersed boundary formulation used herein describes fluid-structure systems in
which an elastic structure is immersed in a viscous incompressible
fluid.  It employs a Lagrangian description of the structural
deformations and the stresses generated by those deformations, and an
Eulerian description of the momentum, incompressibility, and viscosity
of the coupled fluid-structure system.  In the present model, we
employ a fiber-based description of the thin valve leaflets, and we
describe the aortic wall as an incompressible hyperelastic solid.  Let
$\q \in U \subset \RR^2$ indicate Lagrangian curvilinear coordinates
attached to the valve leaflets, with $\q = (q,r)$, let $\X \in V
\subset \RR^3$ indicate the Lagrangian material coordinate system of
the aortic wall, with $\X = (X,Y,Z)$, and let $\x \in \Omega$ indicate
the Eulerian physical coordinates of the physical domain, with $\x =
(x,y,z)$.  The physical position of fiber point $\q$ at time $t$ is
given by $\vphi(\q,t) \in \Omega$, and the physical position of aortic
wall material point $\X$ at time $t$ is given by $\vchi(\X,t) \in
\Omega$.  We use $\Omegal(t) \subset \Omega$ to indicate the
(codimension 1) physical region occupied by the valve leaflets at time
$t$, $\Omegaw(t) \subset \Omega$ to indicate the (codimension 0)
physical region occupied by the solid-body model of the vessel wall,
and $\Omegaf(t) = \Omega \setminus \Omegaw(t)$ to indicate the
physical region occupied by the fluid at time $t$.

The equations of motion for the coupled fluid-structure system are:
\begin{align}
  \rho \frac{{\mathrm D}\u}{{\mathrm D}t}(\x,t) &= - \grad p(\x,t) + \mu \grad^2 \u(\x,t) + \f(\x,t) + \g(\x,t), \label{e:momentum} \\
  \grad \cdot \u(\x,t) &= 0, \label{e:incompressibility} \\
  \f(\x,t) &= \int_U \F(\q,t) \, \delta(\x - \vphi(\q,t)) \, \Dq, \label{e:fiber_force_density} \\
  \g(\x,t) &= \int_V \grad_{\X} \cdot \PP(\X,t) \, \delta(\x - \vchi(\X,t)) \, \DX \nonumber \\
  & \ \ \ \mbox{} - \int_{\p V} \PP(\X,t) \, \N(\X) \, \delta(\x - \vchi(\X,t)) \, \DA, \label{e:solid_force_density} \\
  \D{\vphi}{t}(\q,t) &= \int_\Omega \u(\x,t) \, \delta(\x - \vphi(\q,t)) \, \Dx, \label{e:fiber_interp} \\
  \D{\vchi}{t}(\X,t) &= \int_\Omega \u(\x,t) \, \delta(\x - \vchi(\X,t)) \, \Dx, \label{e:solid_interp}
\end{align}
in which $\rho$ is the mass density, $\mu$ is the dynamic viscosity,
$\u(\x,t)$ is the Eulerian velocity field of the fluid-structure
system, $p(\x,t)$ is the Eulerian pressure field of the
fluid-structure system, $\f(\x,t)$ is the Eulerian elastic force
density generated by deformations to the fiber model of the valve
leaflets, $\g(\x,t)$ is the Eulerian elastic force density generated
by deformations to the solid-body model of the vessel wall, $\F(\q,t)$
is the Lagrangian elastic force density of the fiber model,
$\PP(\X,t)$ is the first Piola-Kirchhoff elastic stress tensor of the
solid model, $\N(\X)$ is the outward unit normal along $\p V$,
$\delta(\x) = \delta(x) \, \delta(y) \, \delta(z)$ is the
three-dimensional Dirac delta function, and $\frac{{\mathrm D}(\cdot)}{{\mathrm D}t} = \D{(\cdot)}{t} + \u \cdot \grad (\cdot)$ is the material derivative.  In the present work, we
assume uniform mass density $\rho$ and dynamic viscosity $\mu$.  These
assumptions are not essential, however, and versions of the IB method
have been developed that permit the use of spatially varying mass
densities \cite{zhu2002, kim2003, kim2007, mori2008, fai2013,
  roy-submitted} and viscosities \cite{fai2013, roy-submitted}.

Eqs.~\eqref{e:momentum} and \eqref{e:incompressibility} are the
Eulerian incompressible Navier-Stokes equations.  Here, the momentum
equation \eqref{e:momentum} is augmented by two Eulerian body force
densities.  The first of these, $\f(\x,t)$, is determined by
eq.~\eqref{e:fiber_force_density} to be the Eulerian force density
that is equivalent to the Lagrangian fiber force density $\F(\q,t)$.
The second body force in \eqref{e:momentum}, $\g(\x,t)$, is determined
by eq.~\eqref{e:solid_force_density} to be the Eulerian force density
equivalent to the Lagrangian description of the forces generated by
the solid-body model of the vessel wall, which are expressed in terms
of $\PP(\X,t)$.  Notice that $\f(\x,t)$ is a singular force density
supported on $\Omegal(t)$.  By contrast, $\g(\x,t)$ is supported on
$\Omegaw(t)$.  Away from $\p\Omegaw(t)$, $\g(\x,t)$ is nonsingular,
although $\g(x,t)$ is singular along $\p\Omegaw(t)$ if $\PP \, \N \neq
\vec{0}$; see eq.~\eqref{e:solid_force_density}.

Eqs.~\eqref{e:fiber_interp} and \eqref{e:solid_interp} determine the
motions of the immersed structures from the Eulerian material velocity
field $\u(\x,t)$.  Because of the presence of viscosity throughout the
fluid-solid system, $\u(\x,t)$ is continuous, and thus
\eqref{e:fiber_interp} and \eqref{e:solid_interp} are equivalent to
\begin{align}
  \D{\vphi}{t}(\q,t) &= \u(\vphi(\q,t),t), \\
  \D{\vchi}{t}(\X,t) &= \u(\vchi(\X,t),t). \label{e:lag_velocity}
\end{align}
Because $\grad \cdot \u(\x,t) = 0$, the immersed solid is
automatically treated as incompressible in this formulation.
Specifically, at least in the continuum formulation, there is no need
to include terms in the elastic strain-energy functional to penalize
compressible deformations.

In practice, we use a standard $C^0$ finite element method to
approximate the deformations and stresses of the vessel wall.  To do
so, it is useful to adopt a weak formulation for the structural
equations, so that
\begin{align}
  \g(\x,t) &= \int_V \G(\X,t) \, \delta(\x - \vchi(\X,t)) \, \DX, \label{e:weak_solid_force_density_1} \\
  \int_V \G(\X,t) \cdot \V(\X) \, \DX &= - \int_V \PP(\X,t) :
  \grad_{\X} \V(\X) \, \DX, \ \forall
  \V(\X), \label{e:weak_solid_force_density_2} \\
  \D{\vchi}{t}(\X,t) &= \U(\X,t), \label{e:weak_solid_velocity_1} \\
  \int_V \U(\X,t) \cdot \V(\X) \, \DX &= \int_V \u(\vchi(\X,t),t) \cdot \V(\X) \, \DX, \label{e:weak_solid_velocity_2}
\end{align}
in which $\G(\X,t)$ is the Lagrangian elastic force density of the
aortic wall, $\U(\X,t)$ is the Lagrangian velocity field of the aortic
wall, and $\V(\X)$ is an arbitrary Lagrangian test function that is
not assumed to vanish on $\p V$.  In the continuum setting,
eqs.~\eqref{e:solid_force_density} and
\eqref{e:weak_solid_force_density_1}--\eqref{e:weak_solid_force_density_2}
are equivalent definitions for $\g(\x,t)$; however, these formulations
lead to different numerical schemes when discretized, and only
eqs.~\eqref{e:weak_solid_force_density_1}--\eqref{e:weak_solid_force_density_2}
lead directly to a standard nodal finite element structural
discretization.  Further, in the continuum equations,
eqs.~\eqref{e:weak_solid_velocity_1}--\eqref{e:weak_solid_velocity_2}
are equivalent to eq.~\eqref{e:solid_interp} and to
eq.~\eqref{e:lag_velocity}.

\subsection{Valve Leaflets}

\subsubsection{Geometry}

\begin{figure}
  \centering
  \begin{subfigure}[t]{0.2\linewidth}
    \centering
    \vspace{0pt}\includegraphics[width=2.5cm]{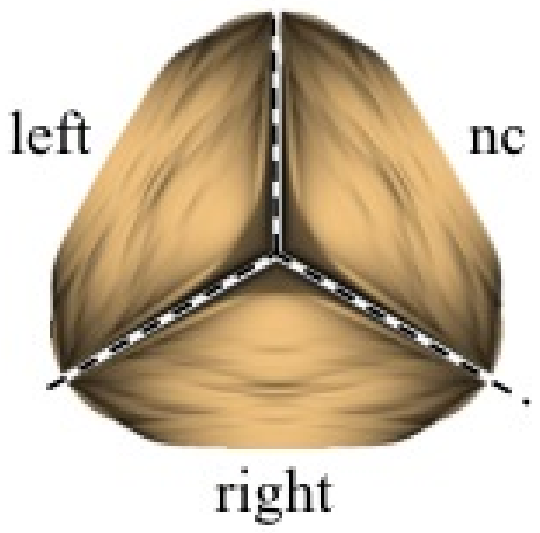}
    \caption{\label{f:aortic_root_modelA}}
  \end{subfigure}
  \hspace{1.5cm}
  \begin{subfigure}[t]{0.2\linewidth}
    \centering
    \vspace{0pt}\includegraphics[width=2.7cm]{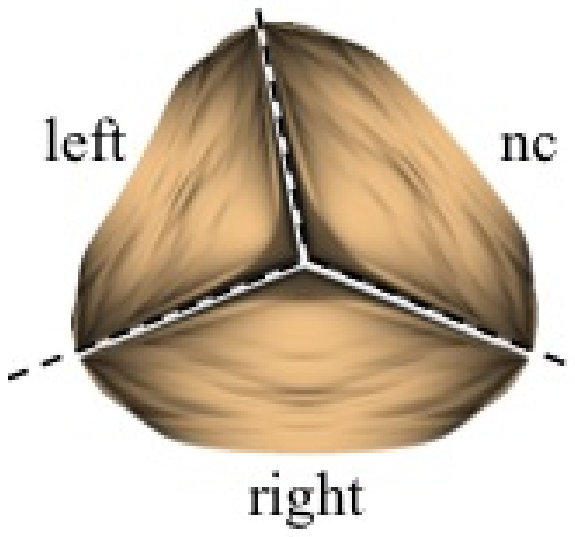}
    \caption{\label{f:aortic_root_modelB}}
  \end{subfigure} \\
  \begin{subfigure}[t]{0.2\linewidth}
    \centering
    \vspace{0pt}\includegraphics[width=2.5cm]{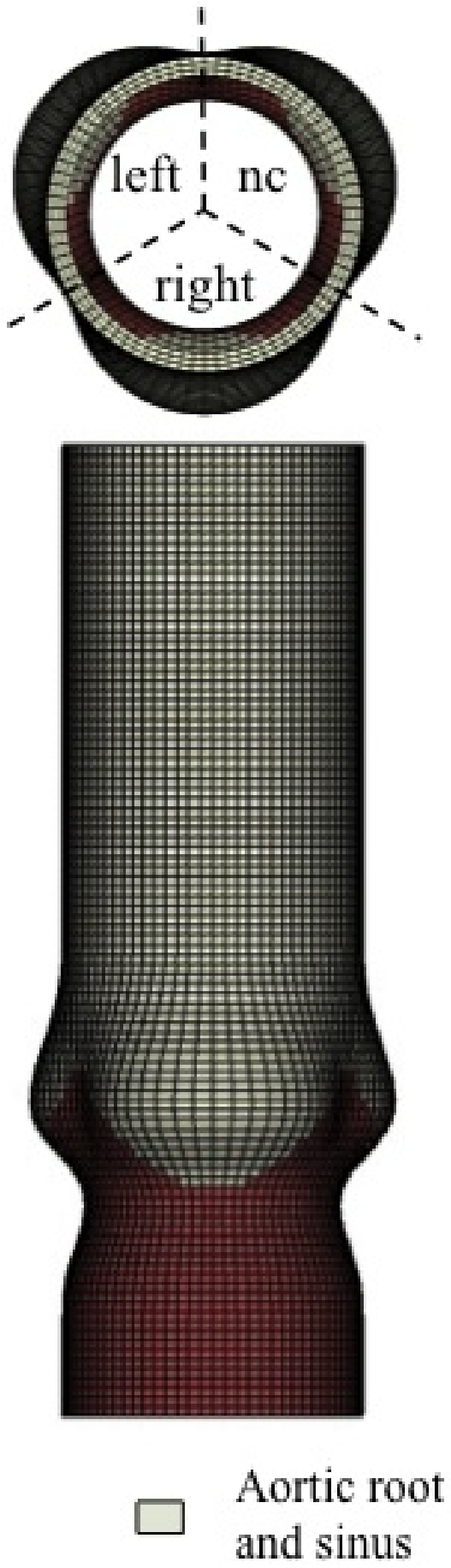}
    \caption{\label{f:aortic_root_modelD}}
  \end{subfigure}
  \hspace{1.5cm}
  \begin{subfigure}[t]{0.2\linewidth}
    \centering
    \vspace{0pt}\includegraphics[width=2.75cm]{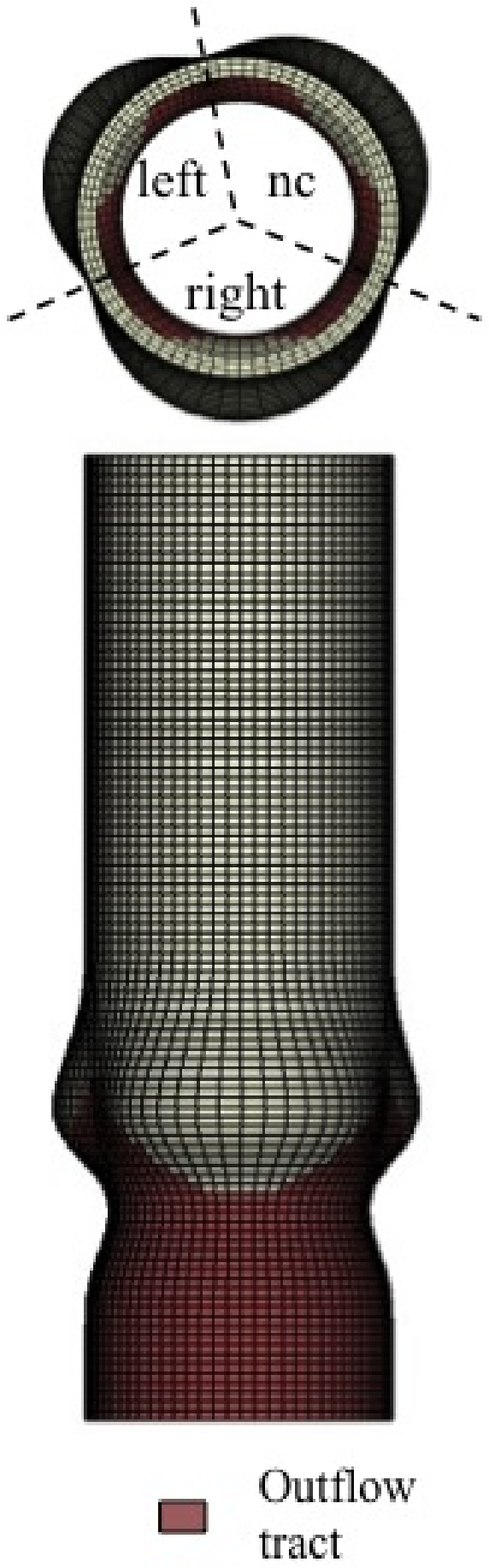}
    \caption{\label{f:aortic_root_model_asym_E}}
  \end{subfigure}
  \caption{Geometrically idealized model of the aortic root. We use
    Peskin and McQueen's theoretical model of the collagen fiber
    architecture of aortic valve leaflets \cite{peskin1994} to
    determine the shape of the leaflets for
    (\subref{f:aortic_root_modelA}) a symmetric and
    (\subref{f:aortic_root_modelB}) an asymmetric aortic valve.  We
    use measurements of human aortic roots \cite{swanson1974,
      reul1990} to determine the geometry of the outflow tract, the
    sinuses, and the ascending aorta. Panel
    (\subref{f:aortic_root_modelD}) shows a side view and a top view
    of the symmetric aortic root model, and panel
    (\subref{f:aortic_root_model_asym_E}) shows a side view and a top
    view of the asymmetric aortic root model.}
  \label{f:aortic_root_model}
\end{figure}

As in earlier work \cite{griffith2009,griffith2012}, the leaflet
geometry is determined by the mathematical theory of the fiber
architecture of aortic valve leaflets developed by Peskin and McQueen
\cite{peskin1994}.  This theoretical model of valve geometry and fiber
architecture derives the shape of the leaflets from their function,
which is to support a uniform pressure load during diastole, when the
valve is closed.  Each valve leaflet is composed of two families of
fibers that are orthogonal to each other.  One family of fibers runs
from commissure to commissure, and the other runs from the bottom
scalloped edge of the aortic sinus to the free edge of the valve
leaflet.  The leaflets are defined in a closed and loaded
configuration; see fig.~\ref{f:aortic_root_model}.  In this
configuration, the radius of each leaflet, measured from the tip of
the valve to the end of the belly region, is 1.45 cm, and the coapting
portion of each leaflet is 0.97 cm tall. The scale of the leaflets is
based on measurements of human aortic roots \cite{swanson1974,
  reul1990}.  The height of the coapting portion of the leaflet is
chosen to be similar to that of the real valve while enabling the
model valve to support a physiological pressure load when closed.

\subsubsection{Mechanical response}

The mechanical behavior of the leaflets is described by a
strain-energy functional $E=E[\vphi(\cdot,t)]$ described previously
\cite{griffith2009,griffith2012}.  Briefly, let the curvilinear
coordinates $\q = (q,r)$ be chosen so that a fixed value of $q$ labels
an individual fiber, and so that $r$ runs along the fibers.
Consequently, $(\p\vphi/\p r) / \| \p\vphi/\p r \|$ is the unit fiber
tangent vector.  The total elastic energy $E$ is the sum of a
stretching energy $E_\text{s}$ and a bending energy $E_\text{b}$,
\begin{align}
  E &=E_\text{s}+E_\text{b}, \label{e:fiber_energy}\\
  E_\text{s}&=\int_{U} \mathcal{E}_\text{s} \left( \left| \D{\vphi}{r} \right|; \q \right) \Dq, \label{e:fiber_stretching_energy}\\
  E_\text{b} &= \frac{1}{2}\int_{U} c_\text{b}(\q) \left| \DD{\vphi}{r} - \DD{\bar{\vphi}}{r} \right|^{2} \Dq. \label{e:fiber_bending_energy}
\end{align}
Eq.~\eqref{e:fiber_stretching_energy} accounts for the total
stretching energy associated with the fibers, in which
$\mathcal{E}_\text{s}$ is a spatially inhomogeneous local stretching
energy with a quadratic length-tension relationship, and
eq.~\eqref{e:fiber_bending_energy} accounts for the total bending
energy associated with the fibers, in which $c_\text{b}$ is a
spatially inhomogeneous bending stiffness and $\bar{\vphi}$ is the
reference configuration, which is taken to be the initial
configuration.  Because the valve leaflets are modeled as thin elastic
surfaces, the bending resistant energy allows the model to account for
the thickness of the real valve leaflets.  Larger values of
$c_\text{b}$ model a thick, stenotic valve, whereas smaller values of
$c_\text{b}$ model a thin, flexible valve.  The resulting Lagrangian
body force density is the Fr\'{e}chet derivative of $E$.

Valve leaflet parameters are empirically determined for a fixed
leaflet discretization by choosing the leaflet stiffness so that the
leaflet supports a diastolic load of approximately 80~mmHg with 10\%
strain in the family of fibers running from commissure to commissure.
This leads to a commissural fiber stiffness of 7.5e6~dyne$/$cm.  We
further assume that these commissural fibers are a factor of 10
stiffer than the radial fibers that run orthogonal to the commissural
fibers \cite{Sauren81}.  The bending stiffness is chosen to be
approximately the smallest value that yields valve leaflets that
successfully coapt at the end of systole.  These material parameters
are chosen only to yield a functional valve, and are not represented
as corresponding to those of an actual valve cusp.  Nonetheless, this
model of the valve leaflets does account for key structures of actual
valve cusps, including the prominent fibers that run from commissure
to commissure and the 10-to-1 anisotropy of real valve leaflets
\cite{Sauren81}.  See refs.~\citen{griffith2009} and
\citen{griffith2012} for further details.  In
preliminary studies, a sensitivity analysis was conducted for the
bending stiffness of the aortic leaflets, and it was found that
variations on the order of $\pm$ 25\% in the bending stiffness have
little effect on the physiological predictions of the models (not
shown).

\subsection{Aortic Wall}
\label{{s:aortic_wall}}
\subsubsection{Geometry}

An idealized model of the aortic root is constructed, as in previous
work \cite{griffith2009,griffith2012}.  The dimensions of this model
are based on measurements by Swanson and Clark \cite{swanson1974} of
human aortic roots collected at autopsy, and the geometry of the model
is based on measurements by Reul et al.~\cite{reul1990} from
angiograms.  The diameter of the aortic portion of the model is 3 cm,
whereas the diameter of the sinus region, measured from a commissura
to the center of a sinus, is 3.5 cm. The overall length of the model
is of 10 cm, and the distance between the annulus and the aortic flow
outlet is 7.75 cm.  The thickness is constant throughout the model and
is 2 mm \cite{shunk2001}.  In our simulations, we employ a semi-rigid
model of the left-ventricular outflow tract while allowing for a fully
flexible description of the aortic sinuses and the ascending aorta.
The aortic annulus, which is the scalloped line of attachment between
the valve leaflets and the aortic sinuses, was considered to be the
lower boundary of the sinuses; see fig.~\ref{f:aortic_root_model}. The
aortic model used in this study differs from previous work \cite{griffith2009,griffith2012} in that here we treat
the aortic sinuses and ascending aorta as an incompressible
hyperelastic material, whereas in earlier studies it was treated as a
rigid structure.

In our simulations, we consider both symmetric and asymmetric models
of the aortic root.  In the symmetric model, we assume that the
initial and reference configurations of the aortic root exhibit
three-fold symmetry, so that each sinus and the corresponding leaflet
occupies an angle of 120$^\circ$.  For the asymmetric model, the
symmetric geometry is modified so that each sinus, along with the
corresponding leaflet, covers a different angle.  Based on the work of
Berdajs \cite{berdajs2002}, the average angle covered by each leaflet
was found to be $136.22^{\circ}$ for the right coronary sinus,
$122.48^{\circ}$ for the noncoronary sinus, and $101.3^{\circ}$ for
the left coronary sinus.  Other characteristic dimensions of the
leaflets, such as the radius or the height of the coapting region,
were taken to be the same as the symmetric model; see
fig.~\ref{f:aortic_root_model}.

Because the geometry of the valve leaflets are defined in a closed and
loaded configuration, it is also necessary to specify the geometry of
the vessel wall in a loaded configuration, and to compute the
corresponding unloaded configuration.  This configuration will depend
on both the prescribed loaded configuration and the constitutive
model.  To determine the unloaded configuration, we employ an
iterative method described by Bols et al.~\cite{bols2012} detailed in
sec.~\ref{s:backward_displacement}.

\subsubsection{Mechanical response}

We describe the elasticity of the aortic sinuses and ascending aorta
using a hyperelastic constitutive model fit to biaxial tensile test
data collected by Azadani et al.~\cite{azadani2012} from tissue
samples from human aortic sinuses.  Because the data reported by
Azadani et al.~indicate a nearly isotropic material response (see
fig.~\ref{f:constitutive_model}), in this work we model the material
response of the aortic sinuses and ascending aorta using an isotropic
strain-energy functional $W$ with an exponential stress-strain
relation,
\begin{equation}
  W = \frac{c}{2b} \left[ \exp(b(I_{1}-3)) - 1 \right],
  \label{e:constitutive_model}
\end{equation}
in which $I_1 = I_1(\CC) = \tr(\CC)$ is the first invariant of the
right Cauchy-Green strain tensor $\CC = \FF^T \FF$, and $\FF =
\p\vchi/\p\X$ is the deformation gradient tensor associated with the
deformation mapping $\vchi:(V,t) \mapsto \Omega$.  Material parameters
were obtained by least-squares fits to experimental data using MATLAB
(Mathworks, Inc., Natick, MA, USA).  The first Piola-Kirchhoff stress
tensor $\PP$ is determined from $W$ via
\begin{equation}
  \PP = \D{W}{\FF} - \ps \, \FF^{-T} = c \exp(b(I_1-3)) \, \FF - \ps \, \FF^{-T},
\end{equation}
in which $\ps$ is a structural pressure-like term that is chosen to
improve the accuracy of the stress predictions of the method
\cite{HGao14-iblv_diastole}.  As in earlier work
\cite{HGao14-iblv_diastole}, we compute $\ps$ via
\begin{equation}
  \ps = c \exp(b(I_1-3)) - \betas \log(I_3),
\end{equation}
in which $I_3 = I_3(\CC) = \det(\CC)$ is the third invariant of
$\CC$ and $\betas = \text{2.5e4}~\text{kPa}$. Thus, $\PP = \vec{0}$
for $\FF = \II$, and the structural model provides an energetic
penalty for any compressible deformations.

\subsection{Driving and Loading Conditions}

A left-ventricular pressure waveform adapted from human clinical data
of Murgo et al.~\cite{murgo1980} is prescribed at the upstream inlet
of the left-ventricular outflow tract to drive flow through the model
aortic root, and downstream loading conditions are provided by a
three-element Windkessel model by Stergiopolus et
al.~\cite{stergiopulos1999} fit to the clinical data of Murgo et
al.~\cite{murgo1980}.  As in earlier work
\cite{griffith2009,griffith2012}, zero pressure boundary conditions
are imposed on the boundary of the fluid-filled region exterior to the
aortic root model.  Coupling between the detailed IB model and the
reduced circulation models is also done as in earlier work
\cite{griffith2009,griffith2012}.

\subsection{Numerical Methods}

We use a locally-refined staggered-grid discretization of the Eulerian
equations along with a finite difference-based discretization of the
fiber model of the valve leaflets and a finite element-based
description of the continuum vessel wall model.  Our approach is
similar to the spatially adaptive IB scheme described previously
\cite{griffith2012}, except that here we also employ a finite
element-based description of the aortic wall, using an approach
introduced by Griffith and Luo \cite{griffith-submitted}.

\subsubsection{Eulerian and Lagrangian spatial discretizations}

The physical domain $\Omega$ is discretized using a locally-refined
Cartesian grid, but for simplicity we describe only the uniform-grid
version of this method; details on the adaptively refined
discretization are provided by ref.~\citen{griffith2012}.  Let
$(i,j,k)$ index the Cartesian grid cells, and let $\x_{i,j,k} =
(\mbox{$(i+\half)$} h , \mbox{$(j+\half)$} h , \mbox{$(k+\half)$} h)$
indicate the position of the center of grid cell $(i,j,k)$, in which
$h$ is the Cartesian grid meshwidth and $\dx = h^3$ is the Cartesian
grid cell volume.  The Eulerian velocity field $\u = (u,v,w)$ is
approximated at the center of each face of the Cartesian grid cells in
terms of the velocity component that is normal to that face:
$u$ is approximated at the locations $\x_{i-\half,j,k}$; $v$ is
approximated at the locations $\x_{i,j-\half,k}$; and $w$ is
approximated at $\x_{i,j,k-\half}$.  The Eulerian force densities $\f$
and $\g$ are approximated in the same staggered-grid fashion.  The
Eulerian pressure $p$ is approximated at the centers of the Cartesian
grid cells.

The deformations and forces associated with the valve leaflets are
approximated on a fiber-aligned curvilinear mesh.  Let $l$ index the
nodes of this mesh, let $\vphi_l$ and $\F_l$ indicate the current
position and Lagrangian force density of node $l$, respectively, and
let $\dq_l$ indicate the area fraction (quadrature weight) associated
with node $l$.  $\F_l$ is computed from $\vphi_l$ using a finite
difference approximation to the fiber force density; see
refs.~\citen{griffith2009} and \citen{griffith2012} for details,
including relevant finite difference formulae.

The deformations, stresses, and resultant forces associated with the
aortic wall are approximated on a volumetric Lagrangian mesh.  Let $e$
index the elements of this mesh, with $V_e$ indicating the volume
associated with element $e$, so that $V = \cup_e V_e$, let $m$ index
the mesh nodes, and let $\varphi_m(\X)$ indicate the interpolatory
Lagrangian finite element basis function associated with node $m$.
The structural deformation and elastic force density are approximated
in a standard manner via
\begin{align}
  \vchi(\X,t) &= \sum_m \vchi_m(t) \, \varphi_m(\X) \\
  \G(\X,t)    &= \sum_m \G_m(t)    \, \varphi_m(\X),
\end{align}
in which $\vchi_m(t)$ are the nodal positions and $\G_m(t)$ are the
nodal force densities.  The deformation gradient $\FF$ is computed by
directly differentiating the approximation to $\vchi(\X,t)$, and $\PP$
is evaluated from the approximation to $\FF$.  The nodal values
$\G_m(t)$ are determined so that
\begin{equation}
  \int_V \G(\X,t) \, \varphi_n(\X) \, \DX = -\int_V \PP(\X,t) \, \grad_{\X}
  \varphi_n(\X) \, \DX
\end{equation}
for all $n$.  This leads to a linear system of equations that defines
the nodal force densities in terms of the nodal deformations.  In
practice, these integrals are evaluated element-by-element using
Gaussian quadrature rules that are exact for the left-hand side but
are generally only approximate for the right-hand side.

\subsubsection{Lagrangian-Eulerian interaction}

To couple the Lagrangian and Eulerian discretizations, we employ
approximations to the integral transforms of the continuum equations
that replace the singular Dirac delta function $\delta(\x) =
\mbox{$\delta(x) \, \delta(y) \, \delta(z)$}$ by a regularized delta
function $\delta_h(\x) = \mbox{$\delta_h(x) \, \delta_h(y) \,
  \delta_h(z)$}$. In our computations, we construct the
three-dimensional regularized delta function either by using the
four-point one-dimensional regularized delta function of Peskin
\cite{peskin2002}, or by using a broadened version of this function
that has a spatial extent of 8 meshwidths.  The same regularized delta
function is used for both the thin model of the valve leaflets and the
volumetric model of the vessel wall.

For the leaflet model, we employ a coupling
approach frequently used with the IB method.  The Lagrangian forces
\mbox{$\F = (F^x,F^y,F^z)$} associated with the leaflets are converted
into equivalent Eulerian forces \mbox{$\f = (f^x,f^y,f^z)$} via
\begin{align}
  f^x_{i-\half,j,k} &= \sum_l F_l^x \, \delta_h(\x_{i-\half,j,k} - \vphi_l) \, \dq_l, \\
  f^y_{i,j-\half,k} &= \sum_l F_l^y \, \delta_h(\x_{i,j-\half,k} - \vphi_l) \, \dq_l, \\
  f^z_{i,j,k-\half} &= \sum_l F_l^z \, \delta_h(\x_{i,j,k-\half} - \vphi_l) \, \dq_l.
\end{align}
This amounts to using the trapezoidal rule to discretize the integral
transform \eqref{e:fiber_force_density}.  We employ the compact
notation
\begin{equation}
  \f = \cSl[\vphi] \, \F
\end{equation}
to express this \emph{force-spreading} operation.  The Eulerian
velocity \mbox{$\u = (u,v,w)$} is used to determine the dynamics
of the physical positions \mbox{$\vphi = (\phi^x,\phi^y,\phi^z)$} of
the material points of the valve leaflets via
\begin{align}
  \d{}{t} \phi_l^x &= \sum_{i,j,k} u_{i-\half,j,k} \, \delta_h(\x_{i-\half,j,k} - \vphi_l) \, \dx, \\
  \d{}{t} \phi_l^y &= \sum_{i,j,k} v_{i,j-\half,k} \, \delta_h(\x_{i,j-\half,k} - \vphi_l) \, \dx, \\
  \d{}{t} \phi_l^z &= \sum_{i,j,k} w_{i,j,k-\half} \, \delta_h(\x_{i,j,k-\half} - \vphi_l) \, \dx.
\end{align}
We use the notation
\begin{equation}
  \d{}{t} \vphi = \cRl[\vphi] \, \u
\end{equation}
to express this \emph{velocity-restriction} operation.  Notice that by
construction, the force-spreading and velocity-interpolation operators
are adjoints, i.e., $\cRl[\vphi] = \cSl[\vphi]^*$.

The Lagrangian-Eulerian coupling scheme used for the vessel model is
based on an approach developed by Griffith and Luo
\cite{griffith-submitted}.  This approach is first to construct a
force-spreading operator, and then to determine a velocity-restriction
operator that is the adjoint of the force-spreading operator.
Briefly, for each element $e$, a set of quadrature points $Q(e)$ is
determined, and for each quadrature point $q \in Q(e)$, $\X_q^e$
indicates the reference coordinates of the quadrature point, and
$\omega_q^e$ indicates the quadrature weight associated.  The
Lagrangian force density $\G = (G^x,G^y,G^z)$ is converted into the
equivalent Eulerian force density $\g = (g^x,g^y,g^z)$ by discretizing
eq.~\eqref{e:weak_solid_force_density_1}, with $\delta(\x)$ replaced
by $\delta_h(\x)$, using this quadrature rule, i.e.
\begin{align}
  g^x_{i-\half,j,k} &= \sum_e \sum_{q \in Q(e)} G^x(\X_q^e,t) \, \delta_h(\x_{i-\half,j,k} - \vchi(\X_q^e,t)) \, \omega_q^e, \\
  g^y_{i,j-\half,k} &= \sum_e \sum_{q \in Q(e)} G^y(\X_q^e,t) \, \delta_h(\x_{i,j-\half,k} - \vchi(\X_q^e,t)) \, \omega_q^e, \\
  g^z_{i,j,k-\half} &= \sum_e \sum_{q \in Q(e)} G^z(\X_q^e,t) \, \delta_h(\x_{i,j,k-\half} - \vchi(\X_q^e,t)) \, \omega_q^e.
\end{align}
Notice that these formulae take advantage of the fact that we may
evaluate the approximations to $\G$ and $\vchi$ at arbitrary
Lagrangian locations.  Specifically, we are not restricted to
evaluating $\G$ and $\vchi$ at the nodes of the finite element mesh.
In practice, the quadrature rules are dynamically determined to ensure
that the physical spacing of the quadrature points is on average
one-half of a Cartesian meshwidth, even under the presence of
extremely large structural deformations.  We use the notation
\begin{equation}
  \g = \cSw[\vchi] \, \G
\end{equation}
to denote the force-spreading operator.

To determine the corresponding velocity-restriction operator, we introduce a Lagrangian
velocity field $\U = (U,V,W)$, which is defined in terms of nodal
velocities $\U_m$ via
\begin{equation}
  \U(\X,t) = \sum_m \U_m(t) \, \varphi_m(\X).
\end{equation}
This velocity field is required to satisfy, for all $n$,
\begin{align}
  & \sum_e \sum_{q \in Q(e)} U(\X_q^e,t) \, \varphi_n(\X_q^e) \,
  \omega_q^e \nonumber \\
  & \mbox{} \hspace{0.125\textwidth} = \sum_e \sum_{q \in Q(e)} \left(\sum_{i,j,k} u_{i-\half,j,k}
  \delta_h(\x_{i-\half,j,k} - \vchi(\X_q^e,t)) \, \dx \right)
  \varphi_n(\X_q^e) \, \omega_q^e, \label{e:discrete_solid_velocity_1}
  \\
  & \sum_e \sum_{q \in Q(e)} V(\X_q^e,t) \, \varphi_n(\X_q^e) \,
  \omega_q^e \nonumber \\
  & \mbox{} \hspace{0.125\textwidth} = \sum_e \sum_{q \in Q(e)} \left(\sum_{i,j,k} v_{i,j-\half,k}
  \delta_h(\x_{i,j-\half,k} - \vchi(\X_q^e,t)) \, \dx \right)
  \varphi_n(\X_q^e) \, \omega_q^e, \label{e:discrete_solid_velocity_2}
  \\
  & \sum_e \sum_{q \in Q(e)} W(\X_q^e,t) \, \varphi_n(\X_q^e) \,
  \omega_q^e \nonumber \\
  & \mbox{} \hspace{0.125\textwidth} = \sum_e \sum_{q \in Q(e)} \left(\sum_{i,j,k} w_{i,j,k-\half}
  \delta_h(\x_{i,j,k-\half} - \vchi(\X_q^e,t)) \, \dx \right)
  \varphi_n(\X_q^e) \, \omega_q^e. \label{e:discrete_solid_velocity_3}
\end{align}
This yields a system of linear equations to be solved for the nodal
values of $\U$.  Notice that
eqs.~\eqref{e:discrete_solid_velocity_1}--\eqref{e:discrete_solid_velocity_3}
correspond to a discrete component-wise approximation to
eq.~\eqref{e:weak_solid_velocity_2}.  We use the notation
\begin{equation}
  \d{}{t} \vchi = \U = \cRw[\vchi] \, \u.
\end{equation}
Notice that $\cRw[\vchi]$ is a nonlocal operator in the sense that it
requires the solution of a system of linear equations.  Although not
shown here, it is the case that $\cRw[\vchi] = \cSw[\vchi]^*$.  See
Griffith and Luo \cite{griffith-submitted} for further discussion.

\subsubsection{Time stepping}

The time stepping scheme is similar to that of Griffith and Luo
\cite{griffith-submitted}.  Let $\dt$ indicate the (uniform) time step
size, and let $[t^n, t^{n+1}] = [n \dt , (n+1) \dt]$ indicate the
$n^\text{th}$ time interval.  In this subsection, superscript indices
always indicate the time step number.  The state variables $\u$,
$\vphi$, and $\vchi$ are defined at integer time steps.  The pressure
$p$, which is not a state variable of the system, is defined at
half-integer time steps.  Approximations to the state variables and to
derived quantities such as the Lagrangian forces are also evaluated at
half-integer time steps.

The time stepping scheme proceeds as follows.  First, predicted
intermediate approximations to the structural deformations $\vphi$ and
$\vchi$ at time $t^{n+\half}$ are determined via
\begin{align}
  \frac{\vphi^{n+\half} - \vphi^{n}}{\dt/2} &= \cRl[\vphi^n] \u^{n}, \\
  \frac{\vchi^{n+\half} - \vchi^{n}}{\dt/2} &= \cRw[\vchi^n] \u^{n}.
\end{align}
This is an application of the forward Euler time stepping scheme.
Lagrangian force densities $\F^{n+\half}$ and $\G^{n+\half}$ are
determined from the predicted structure configurations, and these
Lagrangian force densities are spread to the Cartesian grid via
\begin{align}
  \f^{n+\half} &= \cSl[\vphi^{n+\half}] \F^{n+\half}, \\
  \g^{n+\half} &= \cSw[\vchi^{n+\half}] \G^{n+\half}.
\end{align}

Next, we solve the discretized incompressible Navier-Stokes equations
for $\u^{n+1}$ and $p^{n+\half}$ via a Crank-Nicolson Adams-Bashforth
scheme,
\begin{align}
  \rho\left(\frac{\u^{n+1} - \u^{n}}{\dt} + \left[\u \cdot \grad_h
    \u\right]^{(n+\half)}\right) &= -\grad_h p^{n+\half} + \mu \grad_h^2
  \left(\frac{\u^{n+1} + \u^{n}}{2}\right) \nonumber \\
  & \ \ \ \ \ \ \ \ \ \ \mbox{} + \f^{n+\half} + \g^{n+\half}, \label{e:navier_stokes_1} \\
  \grad_h \cdot \u^{n+1} &= 0, \label{e:navier_stokes_2}
\end{align}
with
\begin{equation}
  \left[\u \cdot \grad_h \u\right]^{(n+\half)} =
  \frac{3}{2}\u^{n} \cdot \grad_h \u^{n} -
  \frac{1}{2}\u^{n-1} \cdot \grad_h \u^{n-1}.
\end{equation}
In the uniform grid case, the discrete operators $\grad_h$,
$\grad_h\cdot\mbox{}$, and $\grad_h^2$ are standard second-order
accurate staggered-grid finite difference approximations to the
gradient, divergence, and Laplace operators, respectively
\cite{griffith2009-efficient}.  We compute $\u \cdot \grad_h \u$ via a
staggered-grid version of the piecewise parabolic method (PPM)
\cite{griffith2009-efficient}.  On locally-refined Cartesian grids,
these discrete operators are straightforward extensions of the uniform
grid operators \cite{griffith2012}.  Solving
eqs.~\eqref{e:navier_stokes_1} and \eqref{e:navier_stokes_2} for
$\u^{n+1}$ and $p^{n+\half}$ requires the solution of the
time-dependent Stokes equations.  The Stokes equations are solved
using an iterative Krylov method with a multigrid preconditioner
derived from the classical projection method
\cite{griffith2009-efficient,cai-submitted}.

Finally, we determine approximations to the structural deformations at
the end of the time step via
\begin{align}
  \frac{\vphi^{n+1} - \vphi^{n}}{\dt} &= \cRl[\vphi^{n+\half}] \u^{n+\half}, \\
  \frac{\vchi^{n+1} - \vchi^{n}}{\dt} &= \cRw[\vchi^{n+\half}] \u^{n+\half},
\end{align}
with
\begin{equation}
  \u^{n+\half} = \frac{\u^{n+1} + \u^{n}}{2}.
\end{equation}
This is an application of the explicit midpoint rule to the structural
dynamics.

In the initial time step, we do not have a value for $\u^{n-1}$ as
required to evaluate the Adams-Bashforth approximation to the
convective term.  Consequently, in the initial time step, we employ a
two-step Runge-Kutta scheme.

Within each time step, coupling between the detailed fluid-structure
interaction model and the reduced Windkessel model of the circulation
is done as described previously \cite{griffith2009, griffith2012}.  In
this approach, we perform only a single Stokes solve per time step,
except in the initial time step, when we perform two Stokes solves
within the context of a two-step Runge-Kutta scheme.

\subsubsection{Backward displacement method}
\label{s:backward_displacement}

To determine the unloaded configuration of the vessel wall, we use an
iterative backward displacement method described by Bols et
al.~\cite{bols2012} implemented using a custom MATLAB script that
interfaces with the ABAQUS (Dassault Syst\`{e}mes Simulia Corp.,
Providence, RI, USA) finite element analysis software.  Because the
constitutive model \eqref{e:constitutive_model} is not provided by
ABAQUS, this model was implemented using a \texttt{UHYPER} subroutine.

Let $\x_m$ denote positions of the finite element mesh nodes in the
\emph{prescribed} loaded configuration, let $\X_m^i$ denote the
corresponding nodal positions in the \emph{computed} reference
configuration after $i$ iterations of the backward displacement
algorithm, and let $\vchi_m^i = \vchi_m[\X^i]$ denote nodal positions
in the \emph{computed} loaded configuration when $\X^i$ is used as the
unloaded reference configuration.  Notice that $\x_m$, $\X_m^i$, and
$\vchi_m^i$ are all defined at the nodes of the finite element mesh.

The algorithm is straightforward: First, we initialize $\X^0 := \x$
(i.e., we use the deformed coordinates as an initial guess for the
unloaded configuration).  Then, given $\X^i$, $\X^{i+1}$ is determined
by
\begin{equation}
  \X^{i+1} := \X^i + (\x - \vchi^i).  \label{e:X_update}
\end{equation}
Notice that if we define the displacement from the computed reference
configuration to the deformed coordinates at step $i$ by $\vec{D}^i =
\vchi^i - \X^i$, then \eqref{e:X_update} is equivalent to
\begin{equation}
  \X^{i+1} := \x - \vec{D}^i.
\end{equation}
Thus, the reference coordinates are determined via a \emph{backward}
displacement from the deformed configuration.  If this process
converges, $\vchi^i \rightarrow \x$ and we obtain reference
coordinates $\X$ such that $\vchi[\X] = \x$.

The convergence of this iterative process is assessed in terms of the
discrete $L^2$ norm of the residual $\x - \vchi^i$, i.e.
\begin{equation}
  r^i = \|\x - \vchi[\X^i]\|_2 = \left(\sum_m \|\x_m - \vchi_m^i\|_2^2\right)^{1/2}.
\end{equation}
The routine was judged to have converged once the residual was less
than 0.1\% of the average vessel radius and the change in the residual
between two consecutive iterations was less than 0.015\% of the
average radius.  In general, obtaining convergence may require the use
of damping, or of gradually increasing the loading pressure; however,
in practice, we have found this algorithm to be robust, and neither
damping nor incremental loading were required for the analyses
performed herein.

\subsubsection{Discretization}

In our dynamic fluid-structure interaction simulations, the
computational domain is taken to be a $10~\text{cm} \times
10~\text{cm} \times 10~\text{cm}$ cubic region discretized using a
three-level locally refined grid.  For the purposes of a mesh
convergence study, two different fine grid spacings are used: a
coarser one corresponding to $h = 0.78125~\text{mm}$, which is
equivalent to a uniform $128 \times 128 \times 128$ Eulerian
discretization, and a finer one corresponding to $h =
0.390625~\text{mm}$, which is equivalent to a uniform $256 \times 256
\times 256$ Eulerian discretization.  The curvilinear mesh used to
discretize the valve leaflets has a physical grid spacing of
approximately $h/4$ for the coarser Eulerian discretization, and to
$h/2$ for the finer discretization.  The volumetric mesh used to
discretize the vessel wall uses 23,400 8-node (trilinear) hexahedral
elements for both the symmetric and the asymmetric model.  The average
mesh aspect ratio was $1.93 \pm 0.25$ for the symmetric root model and
$1.94 \pm 0.34$ for the asymmetric one.  A preliminary convergence
study in ABAQUS was performed to ensure mesh convergence of the
structural model for both the backward displacement method and the FE
analysis.  We use the standard four-point delta function of Peskin
\cite{peskin2002} for the coarser simulations, and a broadened 8-point
version of this kernel function for the higher resolution simulations,
so that the physical extent of the regularized delta function is the
same for both spatial resolutions.\footnote{Similar convergence
  studies, although for substantially different applications, are
  described in refs.~\citen{BEGriffith12-agibm} and
  \citen{HGao14-iblv_diastole}.}  Consequently, the valve leaflets have
  the same effective thickness in for both spatial resolutions.  A
  uniform time step size of $\dt = \text{9.94369e-6}~\text{s}$ was
  used with the coarser Eulerian discretization, whereas a uniform
  time step size of $\dt = \text{1.40625e-5}$ was used for the finer
  Eulerian discretization.  These time step sizes are empirically
determined to be within a factor of $\sqrt{2}$ of the largest time
step that satisfies the stability restriction of our explicit time
integration method.

\section{Results}

\subsection{Constitutive Parameters}

\begin{figure}
  \centering
%
%
\begin{psfrags}%
\psfragscanon%
%
\psfrag{s05}[t][t]{\color[rgb]{0,0,0}\setlength{\tabcolsep}{0pt}\begin{tabular}{c}Stretch\end{tabular}}%
\psfrag{s06}[b][b]{\color[rgb]{0,0,0}\setlength{\tabcolsep}{0pt}\begin{tabular}{c}Cauchy stress (kPa)\end{tabular}}%
\psfrag{s10}[][]{\color[rgb]{0,0,0}\setlength{\tabcolsep}{0pt}\begin{tabular}{c} \end{tabular}}%
\psfrag{s11}[][]{\color[rgb]{0,0,0}\setlength{\tabcolsep}{0pt}\begin{tabular}{c} \end{tabular}}%
\psfrag{s12}[l][l]{\color[rgb]{0,0,0}Isotropic model fit}%
\psfrag{s13}[l][l]{\color[rgb]{0,0,0}Axial experimental data}%
\psfrag{s14}[l][l]{\color[rgb]{0,0,0}Circumferential experimental data}%
\psfrag{s15}[l][l]{\color[rgb]{0,0,0}Isotropic model fit}%
%
\psfrag{x01}[t][t]{1}%
\psfrag{x02}[t][t]{1.05}%
\psfrag{x03}[t][t]{1.1}%
\psfrag{x04}[t][t]{1.15}%
\psfrag{x05}[t][t]{1.2}%
%
\psfrag{v01}[r][r]{0}%
\psfrag{v02}[r][r]{20}%
\psfrag{v03}[r][r]{40}%
\psfrag{v04}[r][r]{60}%
\psfrag{v05}[r][r]{80}%
\psfrag{v06}[r][r]{100}%
\psfrag{v07}[r][r]{120}%
\psfrag{v08}[r][r]{140}%
%
\resizebox{1\textwidth}{!}{\includegraphics{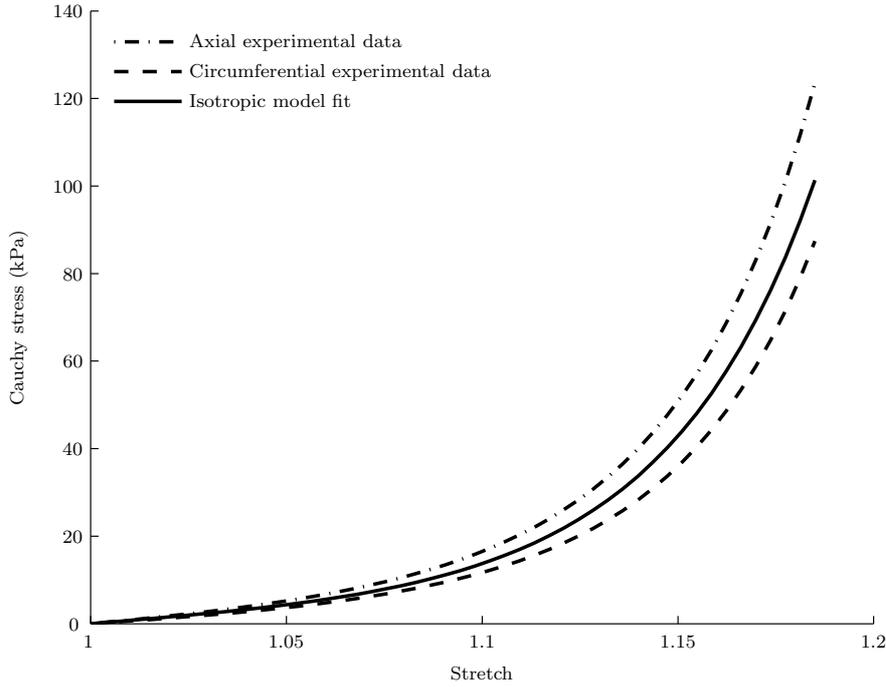}}%
\end{psfrags}%
%

  \caption{Biaxial tensile test data obtained from human aortic root
    tissue samples by Azadani et al.~\cite{azadani2012} and our fit of
    the isotropic hyperelastic constitutive model
    \eqref{e:constitutive_model} to these data.}
  \label{f:constitutive_model}
\end{figure}

The constitutive model \eqref{e:constitutive_model} used to describe
the mechanical response of the aortic sinuses and ascending aorta is
fit to data collected biaxial tensile tests of human aortic root
tissue samples reported by Azadani et al.~\cite{azadani2012}, yielding
model parameter values of $c = 12.8~\text{kPa}$ and $b = 6.9$; see
fig.~\ref{f:constitutive_model}.  Recall that these material
parameters are used in the aortic sinuses and ascending aorta, whereas
the outflow tract is treated as an essentially rigid structure.

\subsection{Unloaded Geometries}

\begin{figure}
  \centering
  \begin{minipage}[b]{0.2\linewidth}
    \centering
    \subcaptionbox{\label{f:loaded_and_unloaded_geometriesA}} {{\includegraphics[height=6cm]{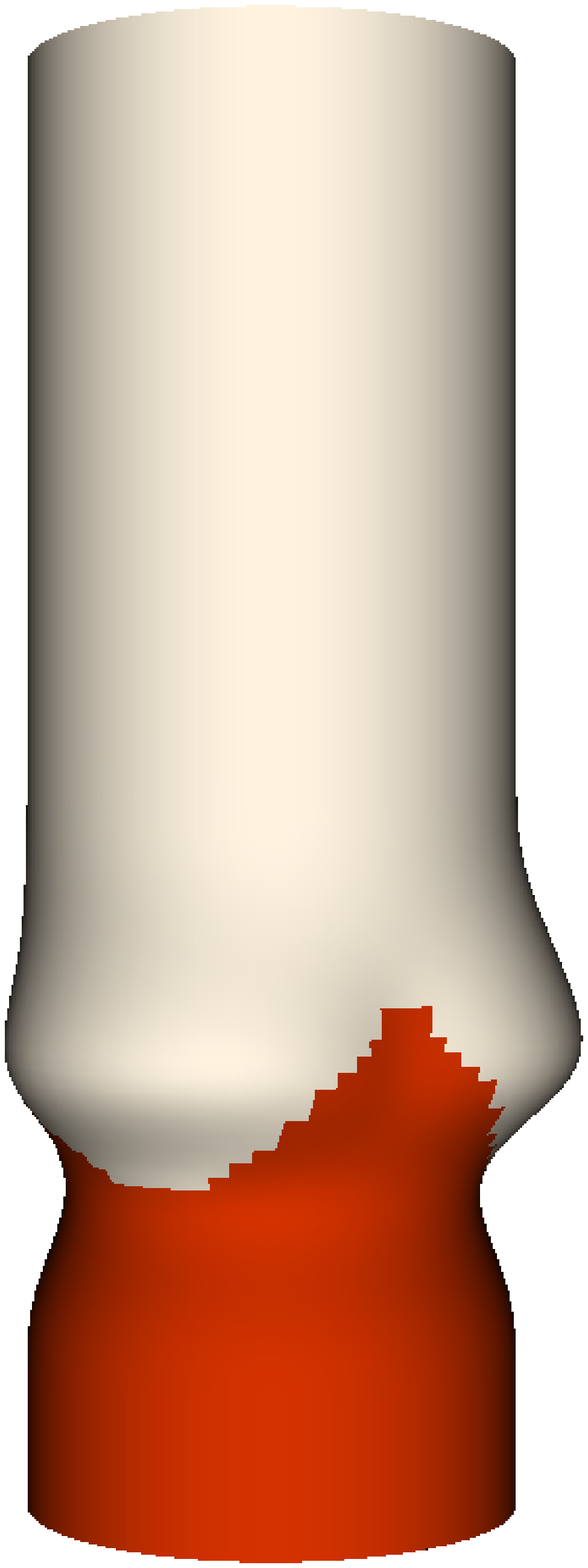}}}
  \end{minipage}
  \begin{minipage}[b]{0.2\linewidth}
    \centering
    \subcaptionbox{\label{f:loaded_and_unloaded_geometriesB}}{\includegraphics[height=6cm]{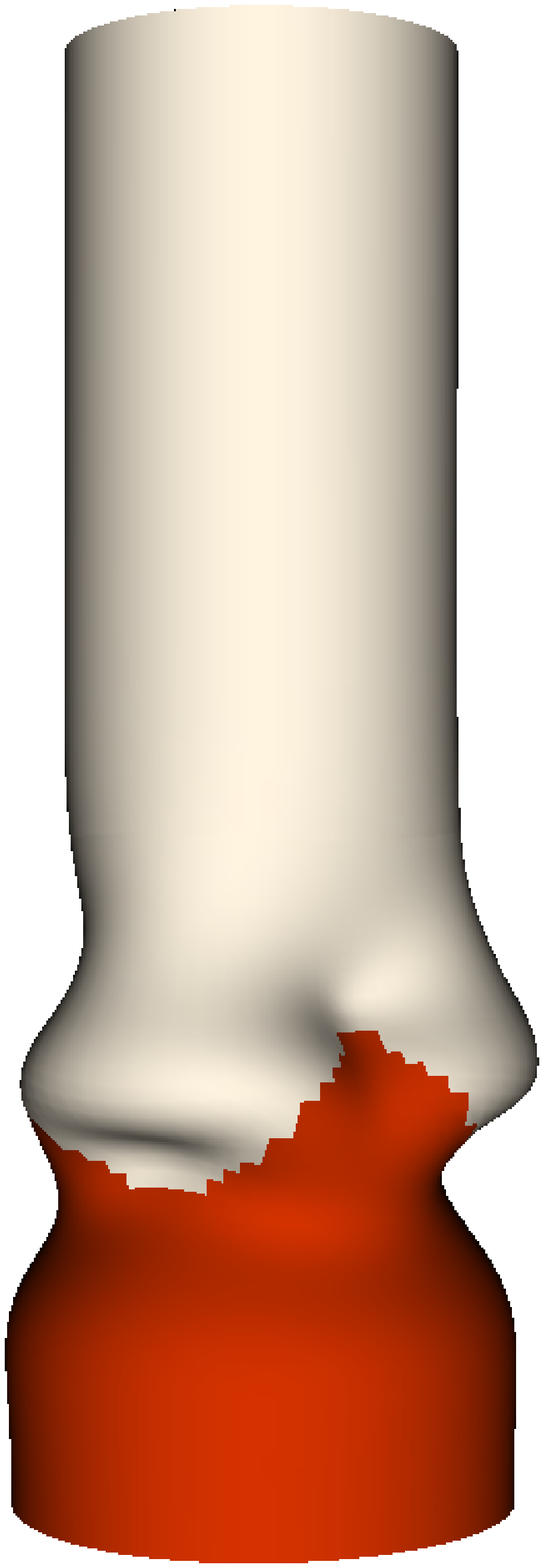}}
  \end{minipage}
  \hspace{0.5cm}
  \begin{minipage}[b]{0.2\linewidth}
    \centering
    \subcaptionbox{\label{f:loaded_and_unloaded_geometriesC}} {{\includegraphics[height=6cm]{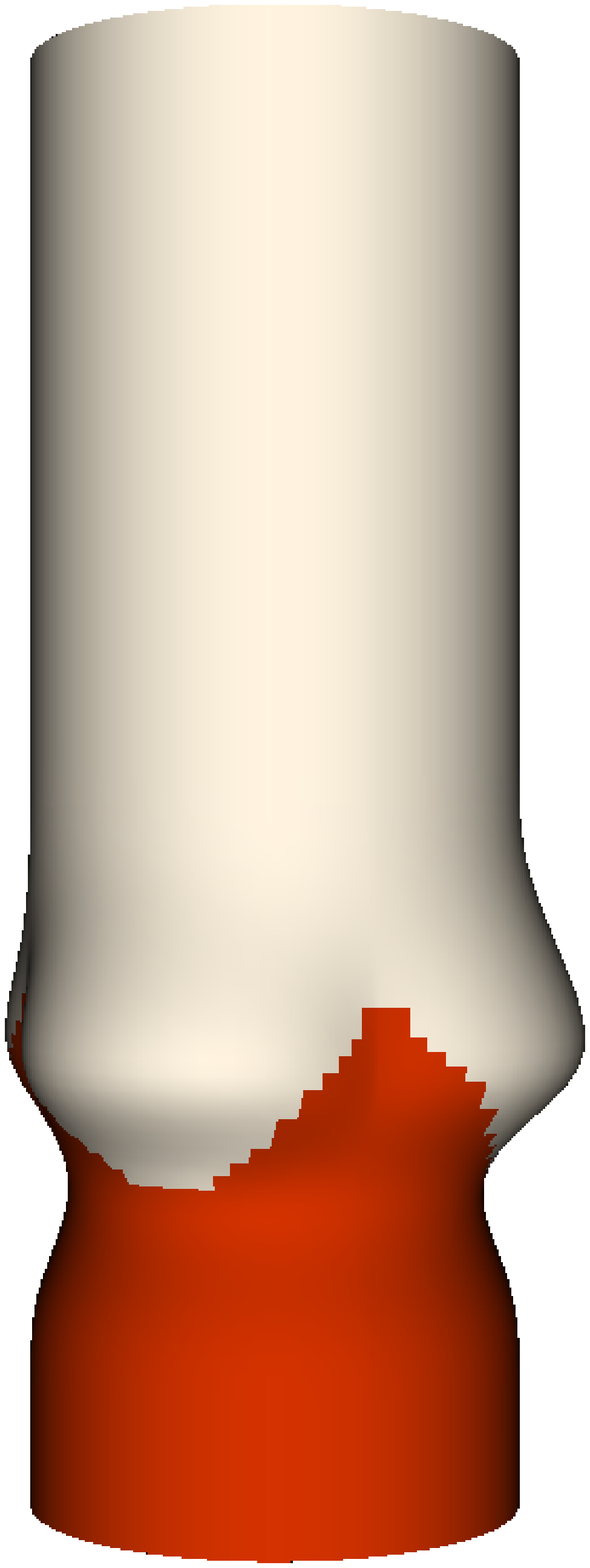}}}
  \end{minipage}
  \begin{minipage}[b]{0.2\linewidth}
    \centering
    \subcaptionbox{\label{f:loaded_and_unloaded_geometriesD}}{\includegraphics[height=6cm]{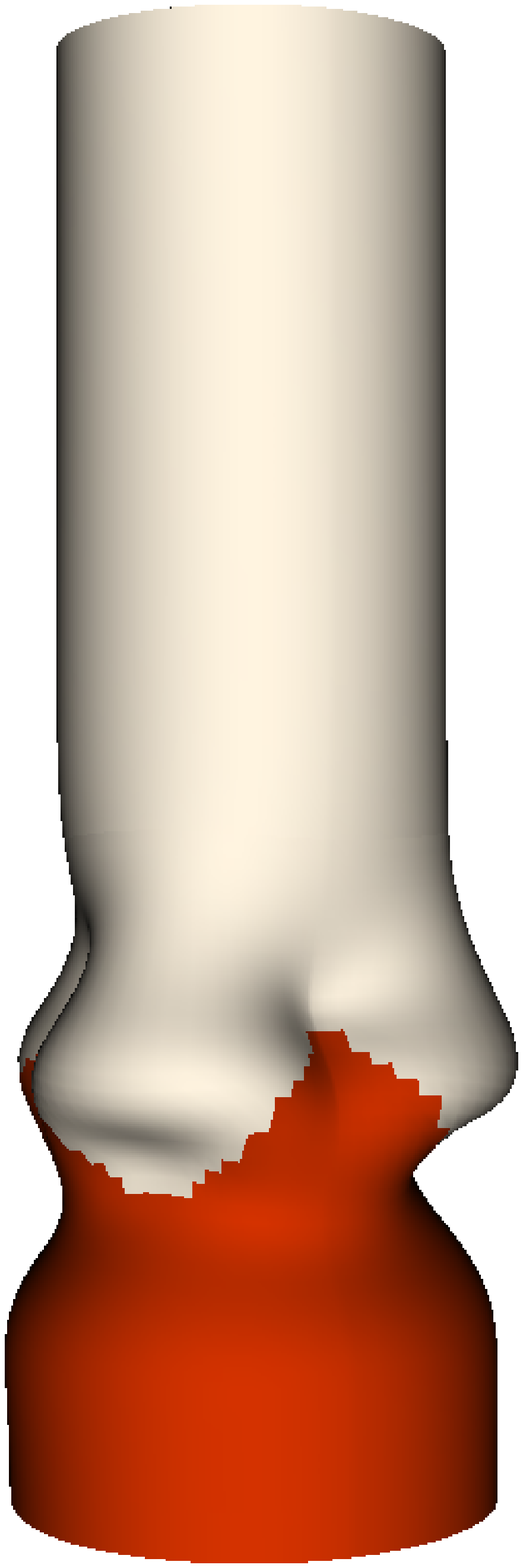}}
  \end{minipage}
  \caption{Views of the (\subref{f:loaded_and_unloaded_geometriesA})
    loaded and (\subref{f:loaded_and_unloaded_geometriesB}) unloaded
    symmetric geometry, and of the
    (\subref{f:loaded_and_unloaded_geometriesC}) loaded and
    (\subref{f:loaded_and_unloaded_geometriesD}) unloaded asymmetric
    geometry.}
  \label{f:loaded_and_unloaded_geometries}
\end{figure}

\begin{figure}
  \centering
%
%
\begin{psfrags}%
\psfragscanon%
%
\psfrag{s05}[b][b]{\color[rgb]{0,0,0}\setlength{\tabcolsep}{0pt}\begin{tabular}{c}residual (m)\end{tabular}}%
\psfrag{s06}[t][t]{\color[rgb]{0,0,0}\setlength{\tabcolsep}{0pt}\begin{tabular}{c}iteration number\end{tabular}}%
\psfrag{s10}[][]{\color[rgb]{0,0,0}\setlength{\tabcolsep}{0pt}\begin{tabular}{c} \end{tabular}}%
\psfrag{s11}[][]{\color[rgb]{0,0,0}\setlength{\tabcolsep}{0pt}\begin{tabular}{c} \end{tabular}}%
\psfrag{s12}[l][l]{\color[rgb]{0,0,0}asymmetric model}%
\psfrag{s13}[l][l]{\color[rgb]{0,0,0}symmetric model}%
\psfrag{s14}[l][l]{\color[rgb]{0,0,0}asymmetric model}%
%
\psfrag{x01}[t][t]{0}%
\psfrag{x02}[t][t]{5}%
\psfrag{x03}[t][t]{10}%
\psfrag{x04}[t][t]{15}%
\psfrag{x05}[t][t]{20}%
\psfrag{x06}[t][t]{25}%
%
\psfrag{v01}[r][r]{$10^{-5}$}%
\psfrag{v02}[r][r]{$10^{-4}$}%
\psfrag{v03}[r][r]{$10^{-3}$}%
\psfrag{v04}[r][r]{$10^{-2}$}%
%
\resizebox{1\textwidth}{!}{\includegraphics{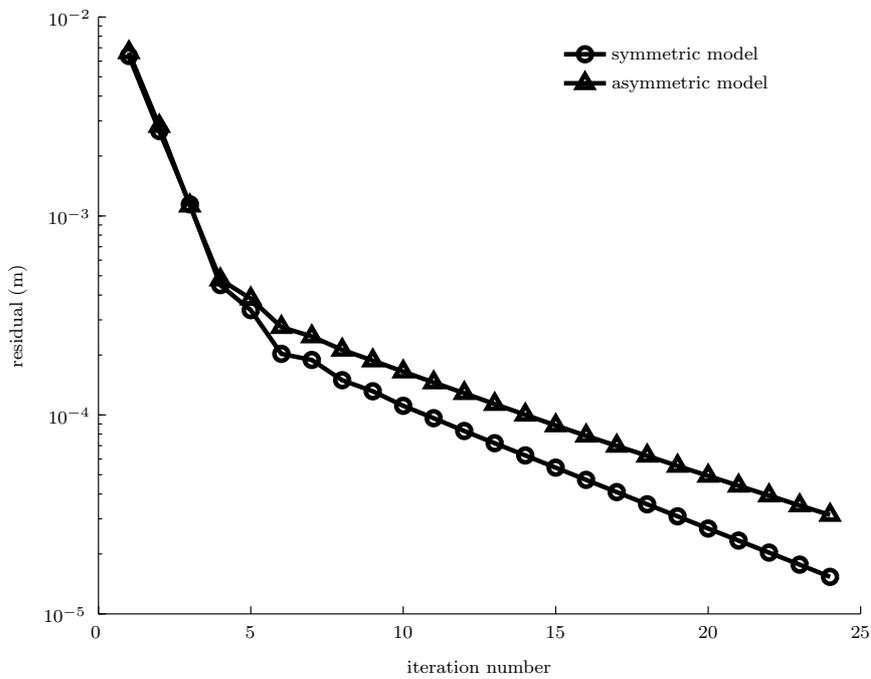}}%
\end{psfrags}%
%

  \caption{Convergence history of the inverse displacement procedure.}
  \label{f:backward_displacement_convergence}
\end{figure}

\begin{figure}
  \centering
  \begin{minipage}[b]{0.1\linewidth}
    \centering
    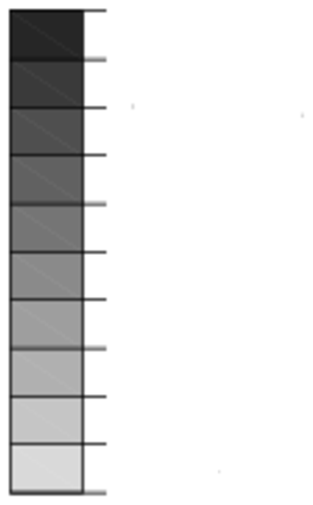
  \end{minipage}
  \hspace{1.0cm}
  \begin{minipage}[b]{0.7\linewidth}
    \centering
    \subcaptionbox{\label{f:backward_displacement_residualA}}
                  {\includegraphics[height=5.8cm]{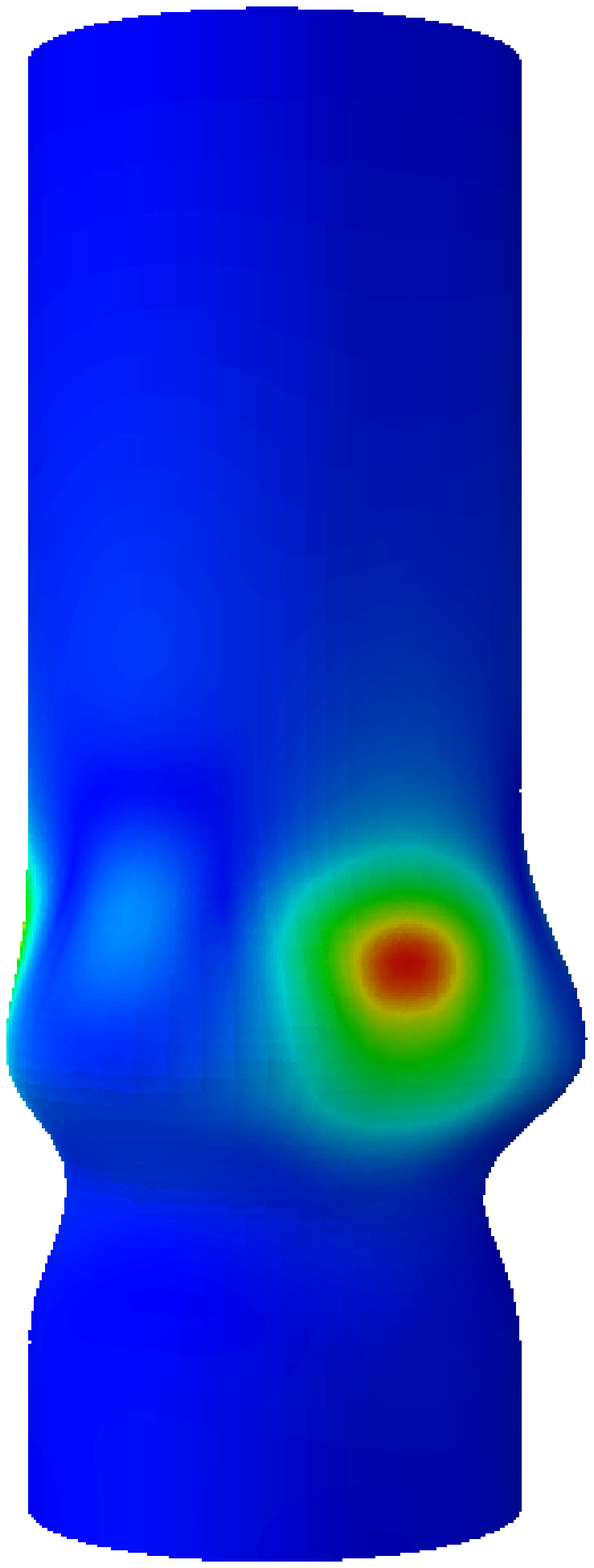}}
                  \hspace{0.5cm}
                  \subcaptionbox{\label{f:backward_displacement_residualB}}    {\includegraphics[height=5.8cm]{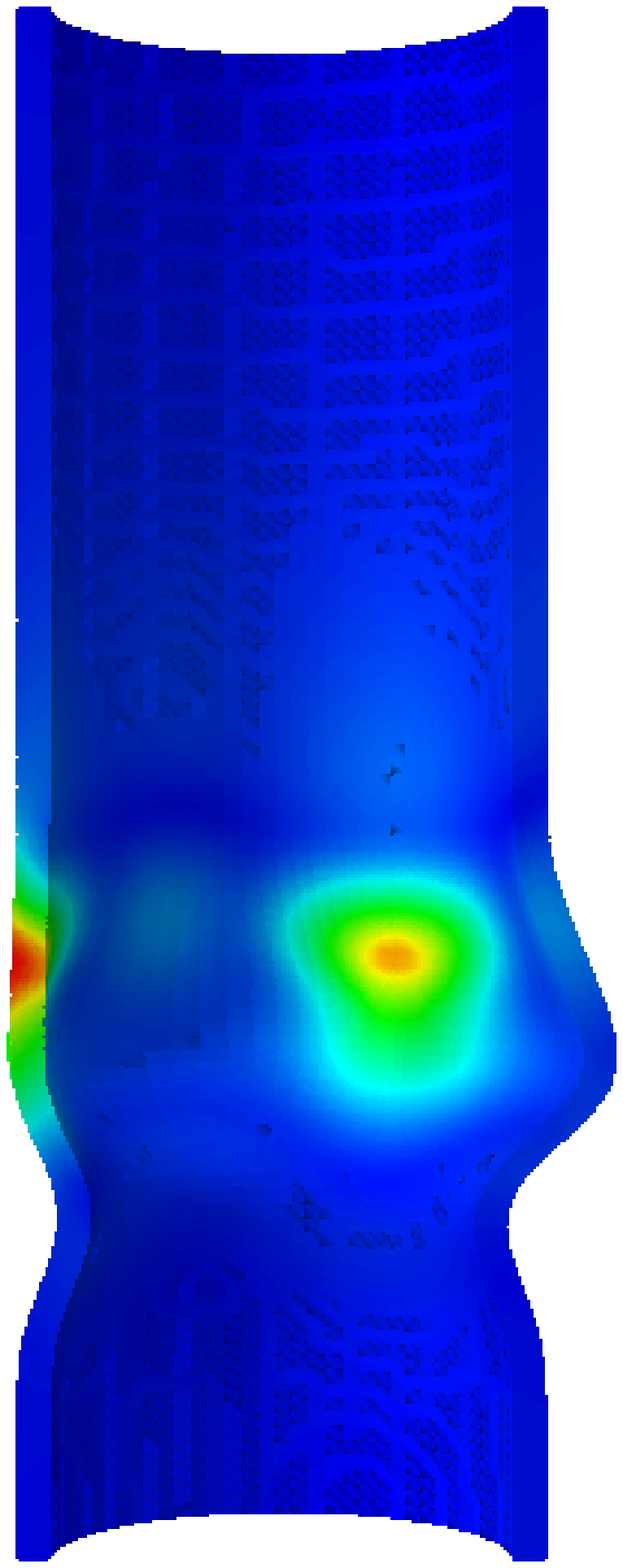}}
  \end{minipage}\\
  \begin{minipage}[b]{\linewidth}
    \centering
    \subcaptionbox{\label{f:backward_displacement_residual_asymA}}
                  {\includegraphics[height=5.8cm]{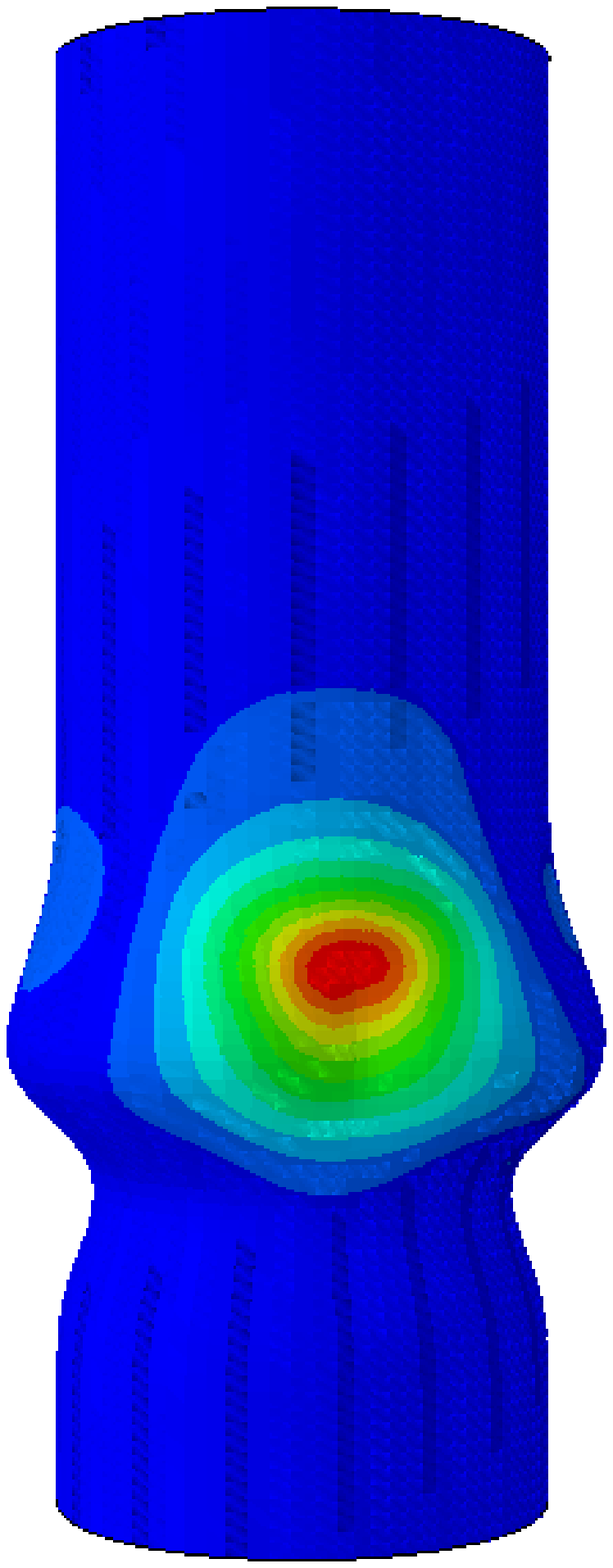} }
                  \hspace{0.5cm}
                  \subcaptionbox{\label{f:backward_displacement_residual_asymB}}    {\includegraphics[height=5.8cm]{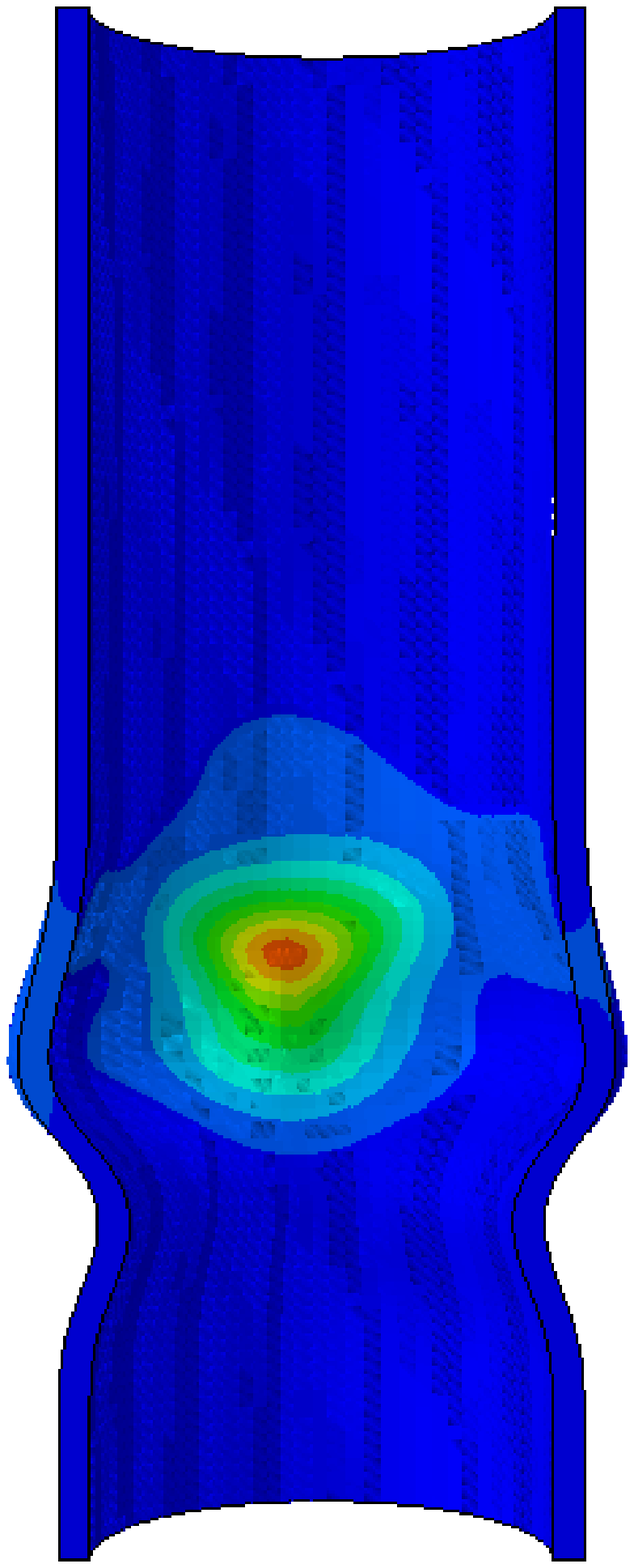} }
                  \hspace{0.5cm}
                  \subcaptionbox{\label{f:backward_displacement_residual_asymC}}
                                {\includegraphics[height=5.8cm]{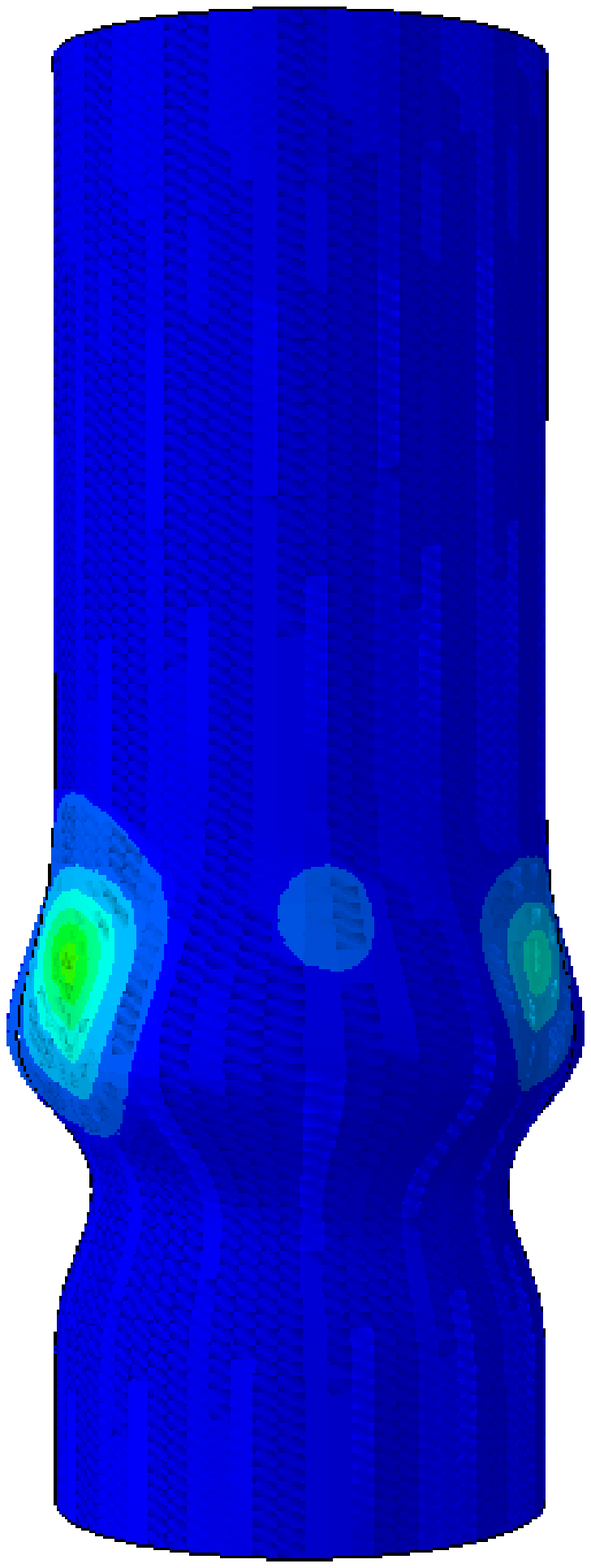} }
                                \hspace{0.5cm}
                                \subcaptionbox{\label{f:backward_displacement_residual_asymD}}    {\includegraphics[height=5.8cm]{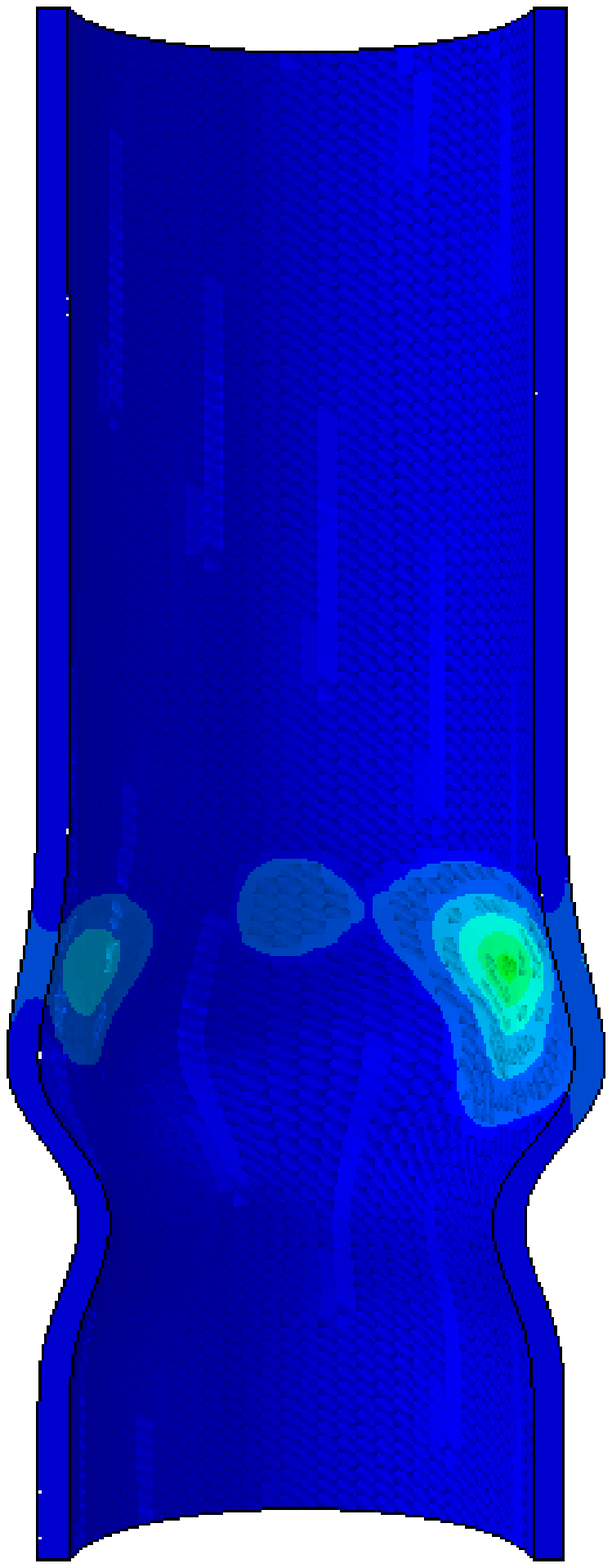}}
  \end{minipage}
  \caption{Contour plots showing the norm of the distance between the
    \emph{prescribed} and \emph{computed} loaded geometries.  Panels
    (a) and (b) show the symmetric geometry, and panels (c)--(f) show
    the asymmetric geometry, with panels
    (\subref{f:backward_displacement_residual_asymA}) and
    (\subref{f:backward_displacement_residual_asymB}) showing the
    right sinus and panels
    (\subref{f:backward_displacement_residual_asymC}) and
    (\subref{f:backward_displacement_residual_asymD}) showing the left
    and noncoronary sinuses.  Recall that the unloaded geometry is
    determined from the loaded geometry.  The computed loaded geometry
    is the result of applying a prescribed pressure load to the
    computed unloaded geometry.}
  \label{f:backward_displacement_residual}
\end{figure}

The unloaded configuration is determined by assuming that the loaded
configuration corresponds to a pressure load of 80 mmHg.  In
fig.~\ref{f:loaded_and_unloaded_geometries}, panels
(\subref{f:loaded_and_unloaded_geometriesA}) and
(\subref{f:loaded_and_unloaded_geometriesB}) show the prescribed
loaded geometry, and panels
(\subref{f:loaded_and_unloaded_geometriesC}) and
(\subref{f:loaded_and_unloaded_geometriesD}) show the computed
zero-pressure configuration.
Fig.~\ref{f:backward_displacement_convergence} shows the convergence
history of the backward displacement method.  When the computed
zero-pressure geometry is subjected to a 80 mmHg pressure load, the
deformed geometry agrees with the prescribed geometry to within
approximately 15 $\mu$m; see
fig.~\ref{f:backward_displacement_residual}.  The discrepancy between
the prescribed loaded geometry and the computed loaded geometry is
greatest at the commissures and smallest along the straight portion of
the vessel.  Similar results are obtained for both the symmetric and
asymmetric models.

\subsection{Fluid-Structure Interaction Analyses}

\subsubsection{Symmetric model}

\begin{figure}
  \input{figure_sym_P_flowrate}
  \caption{Results of a dynamic fluid-structure interaction analysis
    of the symmetric aortic root over the final three simulated cardiac cycles using a
    coarser Cartesian grid (effective grid resolution of $N=128$) and
    a finer Cartesian grid (effective grid resolution of
    $N=256$). (\subref{Fig7A}) Prescribed left-ventricular driving
    pressure and computed aortic loading pressure, and
    (\subref{Fig7B}) computed flow rate, in which the dashed vertical
    lines highlight the moment that the valve leaflets open and the
    solid vertical lines highlight the moment that the valve leaflets
    close.  Large oscillations indicate the reverberations of the
    aortic valve leaflets upon closure.}
  \label{Fig7}
\end{figure}

\begin{figure}
  \input{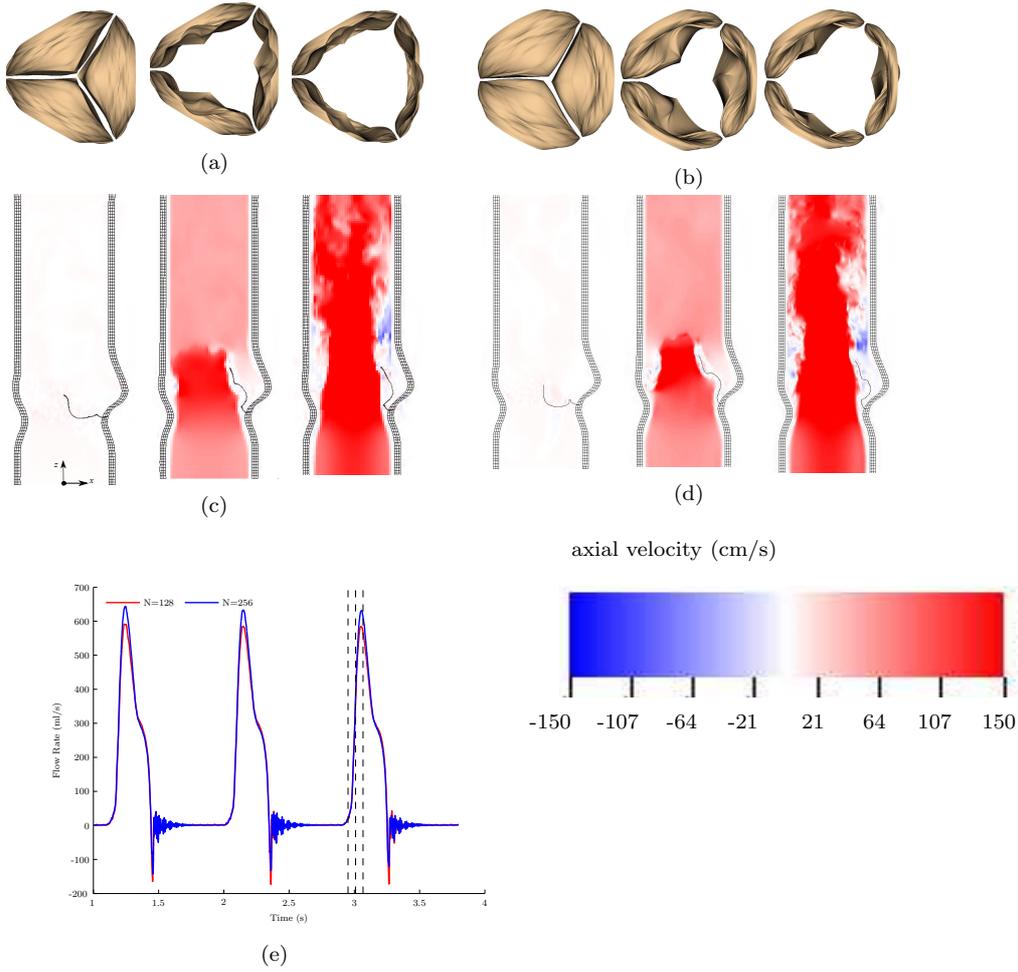}
  \caption{Valve leaflet configurations during valve opening shown at
    equally spaced times during the fourth simulated cardiac cycle for
    (\subref{Fig13A}) the coarser Cartesian grid (effective grid
    resolution of $N=128$) and (\subref{Fig13B}) the finer Cartesian
    grid (effective grid resolution of $N=256$).  Panels
    (\subref{Fig13C}) and (\subref{Fig13D}) show the corresponding
    axial velocity profiles.  Panel (\subref{Fig13E}) shows the
    computed flow rates, with the dashed lines indicating the time
    instants shown in panels (\subref{Fig13A})--(\subref{Fig13D}).}
  \label{Fig13}
\end{figure}

\begin{figure}
  \input{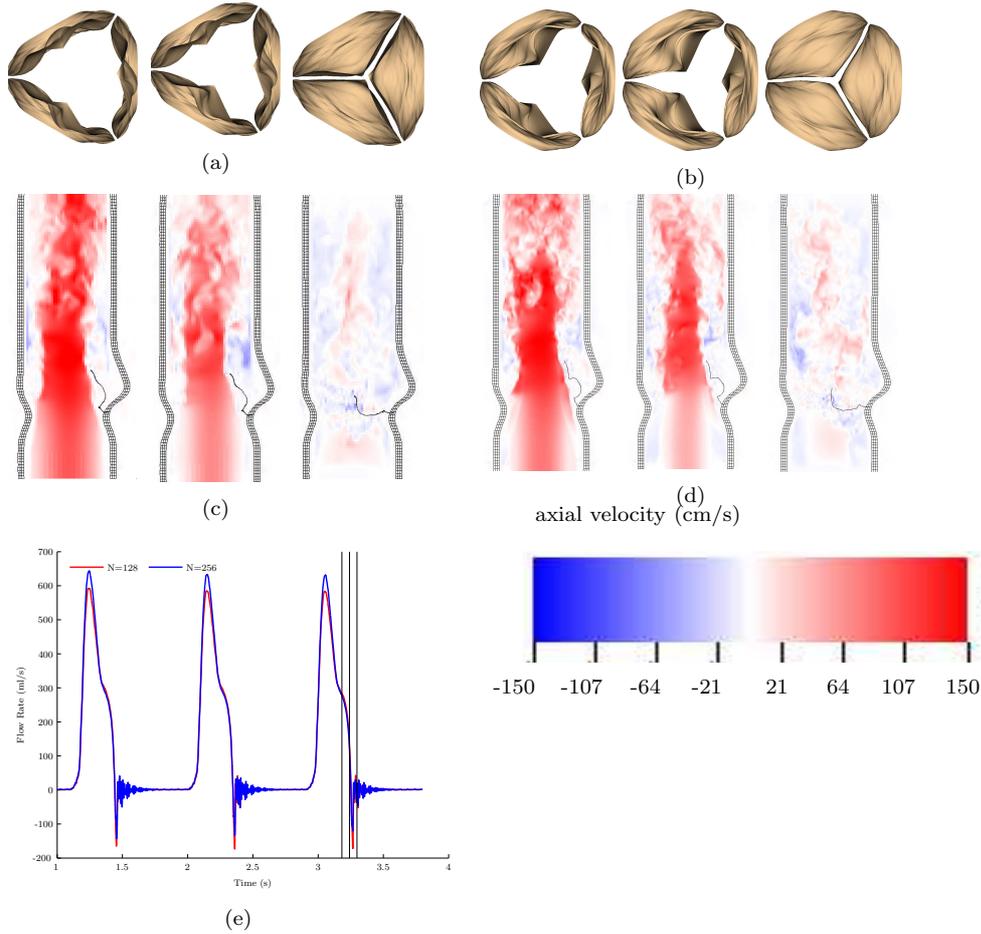}
  \caption{Similar to fig.~\ref{Fig13}, but here showing the dynamics
    of the closure of the aortic valve during the fourth cycle.  In
    this case, the solid lines indicate the instants shown in the
    figure panels.}
  \label{Fig15}
\end{figure}

\begin{figure}
  \input{figure_disp_sym}
  \caption{Displacement contours along the aortic root model.  Panel
    (\subref{Fig14A}) shows displacement contours along the axial
    section of the aortic wall during valve opening, shown as the same
    time instants as in fig.~\ref{Fig13}, for the coarser Cartesian
    grid spacing.  Panel (\subref{Fig14B}) is similar, but here shows
    results for the finer Cartesian grid spacing.  Panels
    (\subref{Fig14C}) and (\subref{Fig14D}) are similar, but here show
    the displacement patterns during valve closure.}
  \label{Fig14}
\end{figure}

\begin{figure}
  \input{figure_disp_sym_plot}
  \caption{Displacements of the deformable aortic root model at four
    locations: (\subref{f:disp_sym_a}) the straight portion of the
    aorta; (\subref{f:disp_sym_b}) just above one of the commissures;
    (\subref{f:disp_sym_c}) the sinotubular junction; and
    (\subref{f:disp_sym_d}) the middle of the sinus. }
  \label{f:disp_symm}
\end{figure}

\begin{figure}
  \input{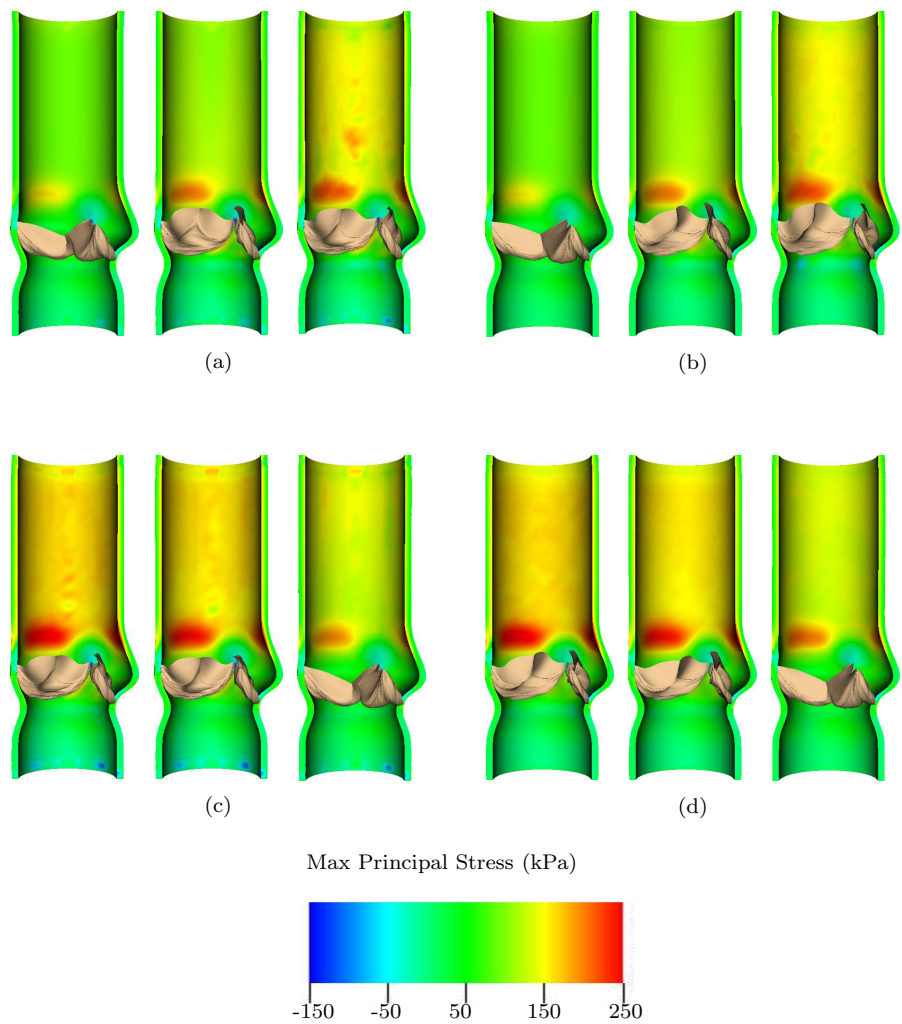}
  \caption{Maximum principal stress contours along the aortic root
    model.  Panel (\subref{f:contour_sp1_a}) shows displacement
    contours along the axial section of the aortic wall during valve
    opening, shown as the same time instants as in fig.~\ref{Fig13},
    for the coarser Cartesian grid spacing.  Panel
    (\subref{f:contour_sp1_b}) is similar, but here shows results for
    the finer Cartesian grid spacing.  Panels
    (\subref{f:contour_sp1_c}) and (\subref{f:contour_sp1_d}) are
    similar, but here show the displacement patterns durning valve
    closure.} \label{f:contour_sp1}
\end{figure}

\begin{figure}
  \input{figure_sp1_sym_plot}
  \caption{Maximum principal stress of the deformable aortic root
    model at four locations: (\subref{f:sp1_sym_a}) the straight
    portion of the aorta; (\subref{f:sp1_sym_b}) just above one of the
    commissures; (\subref{f:sp1_sym_c}) the sinotubular junction; and
    (\subref{f:sp1_sym_d}) the middle of the sinus.}
  \label{f:sp1_symm}
\end{figure}

A dynamic fluid-structure interaction analysis is performed that
includes four cardiac cycles, including an initialization cycle and three cycles in which the model has attained periodic steady state.  (Data are reported only for the beats in which the model has reached steady state.)  Fig.~\ref{Fig7} shows the prescribed
left-ventricular driving pressure along with the computed aortic
loading pressure determined by the coupled Windkessel model and the
computed flow rate through the symmetric aortic root.  We emphasize
that the flow rate is \emph{not} imposed in the model, but rather is
predicted by the fluid-structure interaction analysis.  For the
simulations using the coarser Cartesian spacing,
$h=0.78125~\text{mm}$, stroke volume is 96.2 ml, peak flow rate is
591.5 ml/s, and cardiac output is 6.5 l/min. For the finer Cartesian
grid spacing, $h=0.3906~\text{mm}$, stroke volume is 100.2 ml, peak
flow rate is 643.8 ml/s, and cardiac output is 6.8 l/min.  Both sets
of simulation results are within 15\% of the hemodynamic parameters of
the subject-specific clinical data used to determine the driving
pressure and the Windkessel model parameters \cite{stergiopulos1999,murgo1980}.
Specifically, the driving pressure and the Windkessel model parameters used in this work
were based on clinical measurements from a subject with stroke volume
of 100 ml, peak flow rate of 560 ml/s, and cardiac output of 6.8
l/min, as reported by Murgo et al.~\cite{murgo1980}.  The simulations
also capture the incisura in the flow profile corresponding to the
backward flow generated by the leaflets. The backward flow volume is
1.77 $\pm$ 0.11 ml, or 1.8\% of the stroke volume, for the coarser
grid spacing, and 1.52 $\pm$ 0.059 ml, or 1.5\% of the stroke volume,
for the finer grid spacing.  The net forward flow volume (i.e., into
the aorta) following leaflet coaptation is 0.02 $\pm$ 0.01 ml.  This
forward flow results from interactions between the elastic aortic root
and the dynamic downstream loading pressure.  As the downstream
pressure falls during diastole, the load on the aortic wall and
leaflets decreases and allows the aortic root to gradually push fluid
into the distal aorta.

The model captures the opening and closing dynamics of the valve at
both Cartesian grid spacings.  Fig.~\ref{Fig13} shows the leaflets
being pushed apart along with the formation of the vortices that help
the leaflets to close at the end of systole.  Fig.~\ref{Fig15} shows
that although there is some net flow towards the left ventricle as the
valve closes, once the valve closes, no further flow is allowed
between the aorta and the left ventricle.

Fig.~\ref{Fig14} shows the distribution of the structural
displacements in the aortic root, and fig.~\ref{f:disp_symm} shows
displacement as a function of time at four selected locations: the
center of the sinus; the sinotubular junction; the commissure; and the
ascending aorta. Notice that after the first cardiac cycle, the
displacements are essentially periodic in time. Cartesian grid
resolution has a minimal effect on the range of displacements except
for the displacements measured at the commissure. The largest
amplitude is observed in the ascending aorta ($0.58\pm 0.05$ mm) and
at the sinotubular junction ($0.52\pm0.02$ mm).  A measure of aortic
stiffness is the radial pulsation, i.e., the ratio of the difference
between systolic radius and diastolic radius to the average radius
\cite{nichols2011}, which in the present analysis was approximately
3.9\% for the coarser Cartesian grid spacing and 2.9\% for the finer
Cartesian grid spacing. These values are similar to the value of 2.5\%
reported in refs.~\citen{humphrey2002} and \citen{nichols2011} for the
ascending aorta.  The distribution of the displacements was also in
good agreement with literature data, with the displacements in the
aortic root greater in the sinotubular junction, and at the
commissures than in the sinuses, as reported also by
Lansac~\cite{lansac2002}.

The distribution of the maximum principal stress in the aortic root
follows a distribution similar to that of the displacement, with the
maximum stresses concentrated in the sinotubular junction, just above
the sinuses, as shown in fig.~\ref{f:contour_sp1}. Stress in the
aortic root follows the pulsatile waveform of the pressure; see
fig.~\ref{f:sp1_symm}. Both Cartesian grid spacings yield similar
stress distributions; compare fig.~\ref{f:contour_sp1} panels
(\subref{f:contour_sp1_a}) and (\subref{f:contour_sp1_b}), but
fig.~\ref{f:sp1_symm} shows some differences along the sinotubular
junction and within the sinuses. The maximum principal stress in the
sinuses is 20\% of that observed in the rest of the aortic root.  The
distribution, and the range of the maximum principal stress predicted
by our model, is in agreement with other analyses available in
literature, in particular with the 220 kPa peak of the maximum
principal stress in the sinotubular junction reported by Conti et
al.~\cite{conti2010}.

\subsubsection{Asymmetric model}

\begin{figure}
  \centering
  \input{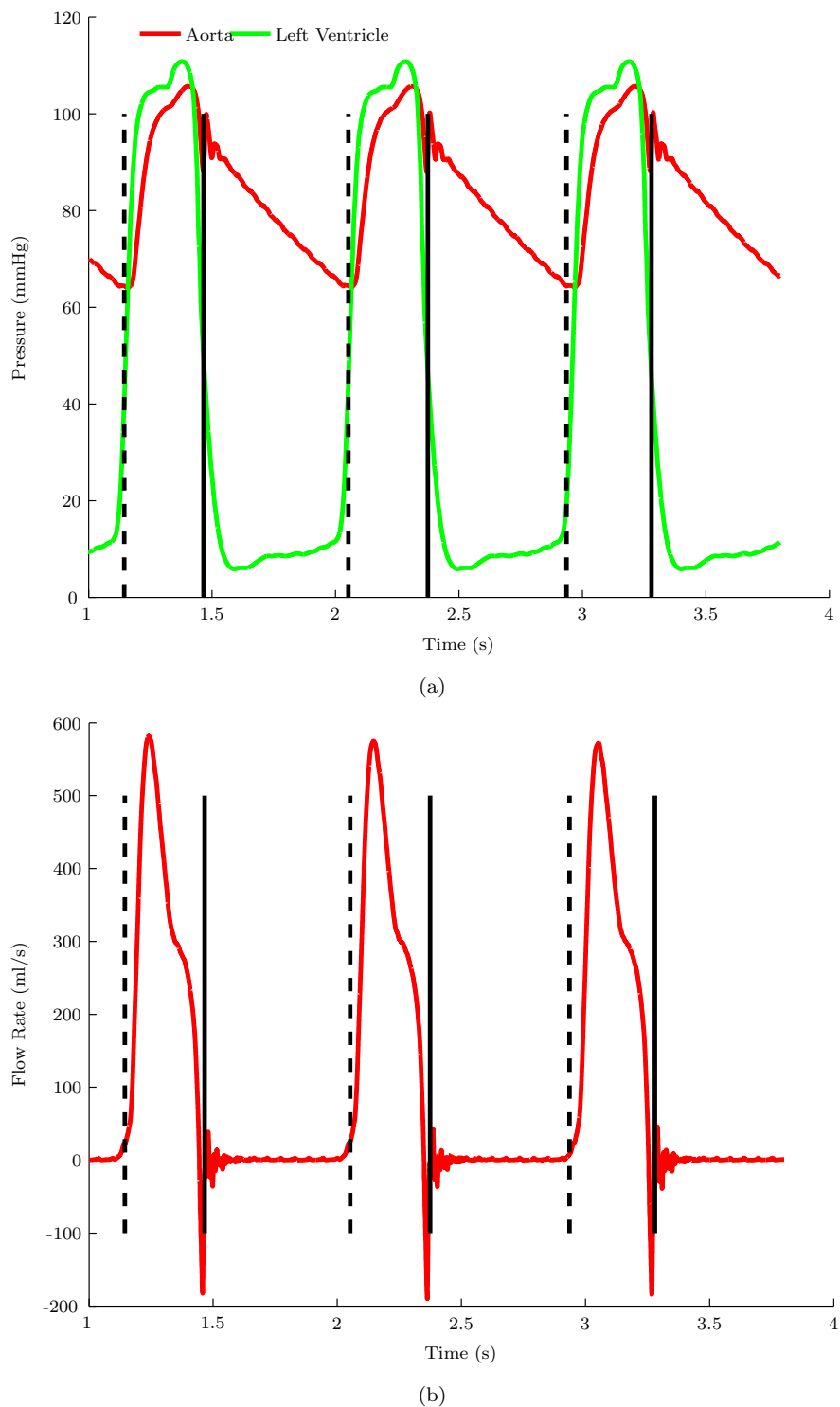}
  \caption{Results of a dynamic fluid-structure interaction analysis
    of the asymmetric aortic root over the final three simulated cardiac
    cycles. (\subref{f:pres_asym}) Prescribed left-ventricular driving
    pressure and computed aortic loading pressure, and
    (\subref{f:flow_asym}) computed flow rate, in which the dashed
    vertical lines highlight the moment that the valve leaflets open
    and the solid vertical lines highlight the moment that the valve
    leaflets close.  Large oscillations indicate the reverberations of
    the aortic valve leaflets upon closure.}
  \label{f:flow_pres_asym}
\end{figure}

\begin{figure}
  \centering
  \input{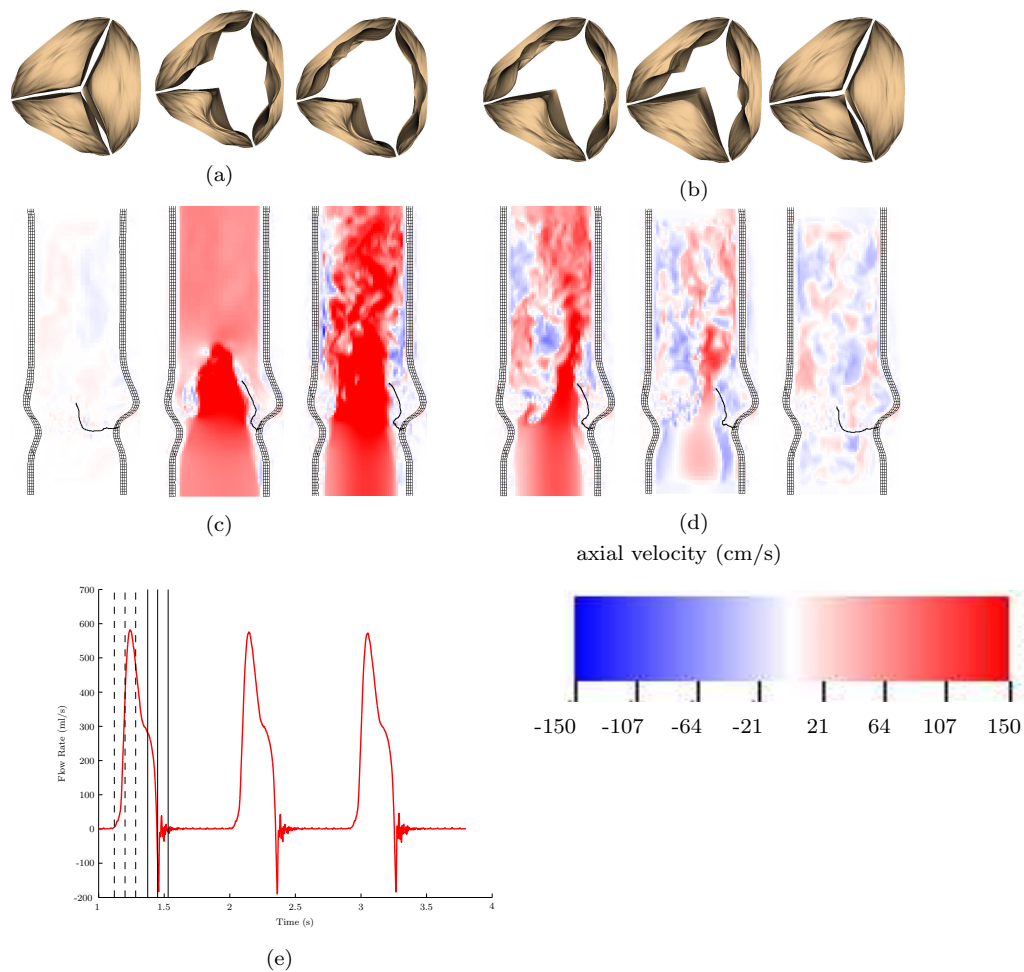}
  \caption{Valve leaflet configurations during
    (\subref{f:asym_leaf_open}) valve opening and
    (\subref{f:asym_leaf_close}) shown at equally spaced times during
    the fourth simulated cardiac cycle.  Panels
    (\subref{f:asym_uz_open}) and (\subref{f:asym_uz_close}) show the
    corresponding axial velocity profiles.  Panel
    (\subref{f:asym_uz_time}) shows the computed flow rates, with the
    dashed lines indicating the times shown during valve opening and
    with the solid lines indicating the times shown during valve
    closure.}
  \label{f:asym_uz}
\end{figure}

\begin{figure}
  \centering
  \input{figure_disp_asym}
  \caption{Displacement contours along the axial section of the
    asymmetric aortic root during valve opening and closure, shown at
    the same time instants as in fig.~\ref{f:asym_uz},
    (\subref{f:asym_dx_open}) during valve opening and
    (\subref{f:asym_dx_open}) during valve closure.}
  \label{f:asym_dx}
\end{figure}

\begin{figure}
  \centering
  \input{figure_disp_nodal_plot}
  \caption{Displacements of the asymmetric aortic root model at four
    locations: (\subref{f:disp_asym_a}) the straight portion of the
    aorta; (\subref{f:disp_asym_b}) just above one of the commissures;
    (\subref{f:disp_asym_c}) the sinotubular junction; and
    (\subref{f:disp_asym_d}) the middle of the sinus.}
  \label{f:disp_asymm}
\end{figure}

\begin{figure}
  \centering
  \input{figure_sp1_contour}
  \caption{Maximum principal stress contours along the axial section
    of the asymmetric aortic root during valve opening and closure,
    shown at the same time instants as in fig.~\ref{f:asym_uz},
    (\subref{f:asym_sp1_open}) during valve opening and
    (\subref{f:asym_sp1_close}) during valve closure.}
  \label{f:asym_sp1}
\end{figure}

\begin{figure}
  \input{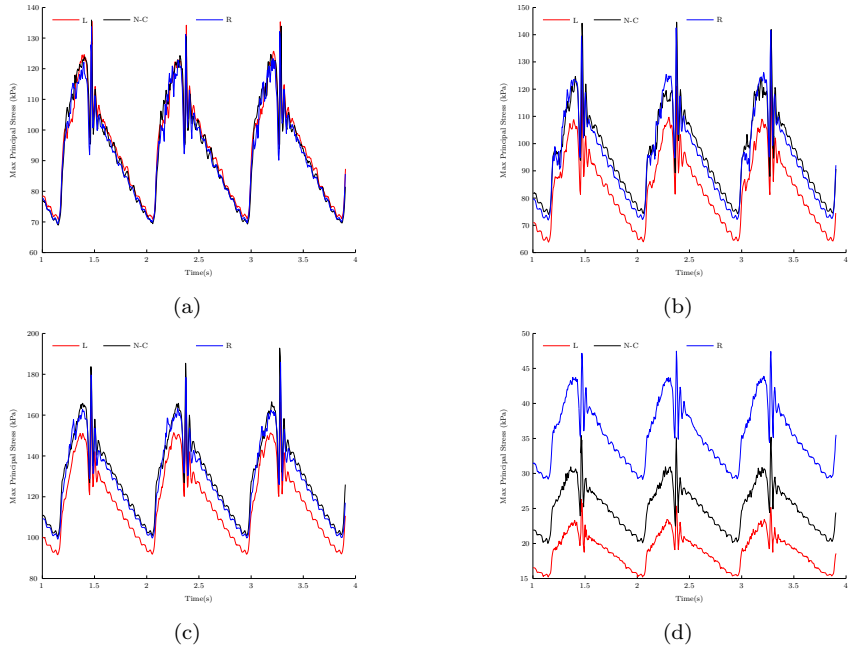}
  \caption{Maximum principal stress of the asymmetric aorta over the
    cycles that reached periodic steady state at four locations:
    (\subref{f:sp1_asym_a}) the straight portion of the aorta;
    (\subref{f:sp1_asym_b}) just above the commissures;
    (\subref{f:sp1_asym_c}) the sinotubular junction; and
    (\subref{f:sp1_asym_d}) the middle of sinus. }
  \label{f:sp1_asymm}
\end{figure}

Similar fluid-structure interaction simulations are performed using
the asymmetric model using only the coarser Cartesian grid spacing of
$h = 0.78125~\text{mm}$. Fig.~\ref{f:flow_pres_asym} shows the
prescribed left ventricular driving pressure and the computed aortic
pressure along with the computed flow rate generated by the
model. Peak flow rate in this model is 581 ml/s, which is reduced
compared to the symmetric studies and is within 5\% of the
subject-specific peak flow rate of 560 ml/s \cite{murgo1980}.  Stroke
volume and cardiac output are 95.1 ml and 6.46 l/min, respectively,
which are both within 6\% of the respective clinical values of 100 ml
and 6.8 l/m \cite{murgo1980}.

The details of the valve kinematics are also different with the
asymmetric model, but the model is still able to capture the leaflets
being pushed back, the flow inversion, the vortices in the sinuses,
and the valve closure, which does not allow regurgitation during
diastole; see figs.~\ref{f:asym_uz}.

Overall, the distribution of the displacements appeared similar to the
symmetric case, with the ascending aorta and the sinotubular junction
moving more than the rest of the aortic root, as shown in
fig.~\ref{f:asym_dx}; however, tracking the displacements over time at
specific locations for the right, left, and noncoronary sinus shows
that the left sinus moves 20\% less than the right sinus and 10\% less
than the noncoronary sinus. In addition, fig.~\ref{f:asym_dx} shows
that the left sector of the aorta maintains a restricted motion
compared to the right and noncoronary sectors of the aorta. This
behavior can be observed also in the aorta in which the radial
pulsation ranges from 1.9\% for the left sector of the aorta, to 3.9\%
for the right sector of the aorta.  The average displacement amplitude
in the sinotubular junction is 0.49 mm for the right and noncoronary
sinus and 0.36 mm for the left sinus. In the ascending aorta, the
average displacement amplitude is 0.54 mm in the right and noncoronary
portion and 0.34 mm in the left portion of the aortic wall.

The global distribution of the maximum principal stress was similar in
the symmetric and asymmetric models, with the maximum stress being
concentrated at the sinotubular junction, in correspondence with the
sinuses; see fig.~\ref{f:asym_dx}. Differences in the maximum
principal stress at the same locations for the right, left, and
noncoronary sinus, are more complex.  Whereas the maximum principal
stress in the aorta is essentially independent of the sector of the
geometry considered, there are large differences in the stress
computed in the left, right, and noncoronary sinus, especially in the
cusp of the sinuses and in the sinotubular junction. Peak maximum
principal stress for the sinotubular junction is 180 kPa, a value
comparable to the 220 kPa value obtained by Conti et
al.~\cite{conti2010}. Peak principal stresses in the sinuses,
evaluated at the cusp, resulted smaller in magnitude than the values
reported by Grande-Allen et al.~\cite{grande1998}, although the
relative differences between the stresses in the sinuses is similar to
those observed in that work.

\section{Discussion}
\label{s:discussion}

In this work, previous IB models of aortic valve dynamics
\cite{griffith2009,griffith2012} are substantially extended by
incorporating a deformable model of the aortic sinuses and ascending
aorta.  This extended model is thereby able to more closely replicate
the complex in vivo dynamics of the real aortic root.  As before, the dynamic
analyses of the aortic valve included four complete cardiac cycles, including a single initialization cycle,
which are shown to be sufficient to yield essentially periodic
results.  As in previous IB models of aortic valve dynamics
\cite{griffith2009,griffith2012}, key hemodynamic parameters,
including aortic pressure, maximum flow rate, cardiac output, and
stroke volume, are in reasonable agreement with physiological ranges
\cite{driscol1965,westerhof2010} and are also in good agreement with
the patient-specific data \cite{murgo1980} used to fit the Windkessel
model \cite{stergiopulos1999} employed in this study.  Further, grid
convergence results demonstrate for the first time that the IB method
is able to yield essentially grid-resolved bulk flow parameters and
vessel deformations at practical grid spacings.

Many strategies have been developed to determine the zero-pressure
geometry of blood vessels
\cite{gee2009,alastrue2010,vavourakis2011,bols2012,dePutter2007,
  sellier2011}. Indeed, medical images of arteries always provide the
vessel geometry in a loaded configuration.  It is known, however, that
using the imaged geometry as the reference geometry of the vessel
underestimates the amount of strain present in the blood vessel
\cite{dePutter2007}.  Among the approaches developed to estimate the
zero-pressure geometry of blood vessels, we chose an iterative
backward displacement approach, as described by Bols et
al.~\cite{bols2012} and by Sellier \cite{sellier2011}.  Iterative
approaches based on geometric considerations are straightforward to
implement and can be applied to a variety of problems and solvers
because they require minimal implementation effort.  In this work, the
backward displacement analysis was implemented in ABAQUS and converged
to a zero-pressure geometry that, when loaded, yielded a deformed
configuration that is in good agreement with the initially prescribed
loaded geometry; see
figs.~\ref{f:loaded_and_unloaded_geometries}--\ref{f:backward_displacement_residual}.
However, the deflated geometry is quite different from the
configuration assumed by a deflated aortic root harvested from a body;
see fig.~\ref{f:loaded_and_unloaded_geometries}. This is a drawback of
these methods, as shown also in a recent study by Wittek et
al.~\cite{wittek2013}, in which the unloaded configuration of the
aorta is different from an unloaded aorta harvested from a body. This
mismatch could be the result of many factors, such as the choice of
the boundary conditions or the need for adding the residual stress to
balance the level of stress within the aortic wall \cite{fung1991}. We
expect that the addition of residual stress estimates, planned in
future work, will reduce the mismatch between the deflated geometry
computed and the deflated aortic root observed in vivo.

Interestingly, the addition of a compliant aortic wall did not
substantially alter the global hemodynamic parameters previously determined by a
noncompliant IB aortic root model \cite{griffith2012}.  In general, we
observed a slight increase in the maximum flow rate as compared to the
noncompliant model, whereas changes in other global flow parameters
were relatively minor.  This finding might suggest that our vessel
wall model is overly stiff despite our use of a constitutive model fit
to biaxial tensile test data from human aortic tissues samples.  In
contrast, we observed that using a compliant aorta affected more
clearly the dynamics of valve leaflet
closure. Figs.~\ref{Fig13}(\subref{Fig13B}) and
\ref{Fig15}(\subref{Fig15B}) show the flow inversion and the
distension in the sinus, which collects fluid during systolic
ejection, and the elastic recoil of the sinus, which pushes down on
the leaflet during closure.  The addition of this mechanism made the
model substantially more robust.  Specifically, we found that the
rigid aortic root model required the valve leaflet properties to be
very carefully tuned in order for the model to remain competent
throughout the cardiac cycle.  This is in clear contrast to the
present compliant aortic root model, which yields a competent valve
for a relatively broad range of leaflet parameters.

Similar model predictions were obtained for both Cartesian grid
spacings used in this study.  We observed hemodynamic values that were
within 5\% of the clinical values reported by Murgo et
al.~\cite{murgo1980} except for the peak flow rate, which was within
5\% of the values reported by Murgo et al.~on the coarser Cartesian
grid and within 15\% on the finer grid.  Cardiac output and stroke
volume were both in very good agreement with the clinical data, within
4\% on the coarser Cartesian grid, and within 1\% on the finer grid.
In addition, the pulsatile radius, a quantity used to measure aortic
stiffness, was also shown to be in relatively good agreement with
literature values in both cases \cite{humphrey2002}.

To make the description of the aortic root dynamics more realistic, we
modified the aortic geometry and the leaflets configuration to
replicate the physiological asymmetry of the sinuses and the
leaflets. Previous work by Grande-Allen et al.~\cite{grande1998} and
by Conti et al.~\cite{conti2010} showed that the use of a realistic
geometry in FE and computational fluid dynamics simulations of the
aortic root yields more accurate results for the stress distribution
along the sinuses.  We based our asymmetric aortic root geometries on
the inter-commissures distances reported by Berdajs
\cite{berdajs2002}.  Overall, the bulk flow parameters were closer to
the original values reported by Murgo \cite{murgo1980} than the
corresponding values obtained with a symmetric model for the same
level of grid refinement.  These results suggest that the use of a
more realistic geometry could yield more accurate estimates of key
hemodynamic quantities.  Structural stresses generated by the immersed
boundary model were also in good agreement with the stresses reported
by Conti et al.~\cite{conti2010}.  Moreover, the level of stress and
the displacement of the right and in the noncoronary sectors of the
aortic root, but not in the left sector of the aortic root, were
similar in the symmetric and the asymmetric case; see
figs.~\ref{f:disp_symm} and \ref{f:disp_asymm}.

Leaflet dynamics for the symmetric aortic valve are shown in
figs.~\ref{Fig13} and \ref{Fig15}, and leaflet dynamics for the
asymmetric aortic valve are shown instead in fig.~\ref{f:asym_uz}.  It
is possible to see, however, that in the asymmetric model, the left
coronary leaflet starts closing before the other two leaflets, thus
bending the flow profile in the aortic tract; see
fig.~\ref{f:asym_uz}(\subref{f:asym_uz_close}).  Because the aortic
root model is asymmetric, there is no reason to expect that the
dynamics of the three leaflets will be identical.  We plan to study
this asymmetry in the leaflet kinematics more carefully in the context
of a more geometrically realistic model of the aortic root that
accounts for the curvature of the real aortic root and ascending
aorta.

\subsection{Limitations and Future Work}

Work is underway to address some of the limitations of these models.
As described previously, our present models do not account for the
curved geometry of the aortic root, or for the residual stresses
present in real arterial vessels.  We specifically aim to incorporate
a description of residual stresses and realistic medical
imaging-derived anatomic geometries into our IB models.  Further, the
present models use an isotropic model of the aortic sinuses and
ascending aorta.  Although the experimental data that were used to fit
this model show an essentially isotropic material response, it is
known that real aortic tissues have a well-defined collagen fiber
structure with an anisotropic material response.  We also aim to
incorporate a more realistic hyperelastic model of the aortic wall
that accounts for families of collagen fibers \cite{gasser2006} as
well as experimentally constrained models of the elasticity of the
aortic valve leaflets \cite{may-newman-2009}.  Further, it is now well
known that the valve leaflets are in fact layered structures
\cite{sacks2007}.  These details are not accounted for in the present
model, and we anticipate that including them in future work will
require an important extension of the methods presented in this study.
Specifically, because the valve leaflets are thin elastic structures,
we expect that using realistic hyperelastic continuum models to
describe the mechanics of the valve leaflets will require adopting a
nonlinear shell formulation.  To our knowledge, continuum shell models
have not yet been widely used within the framework of the IB method,
and we anticipate additional computational research will be required
to implement these types of formulations into our IB simulation
framework.

Another limitation of this work is the long run times required by
these simulations.  Because of the small time step sizes required by
the present algorithm, each cardiac cycle required 50,000 or more time
steps, requiring on the order of one day per cycle on 32--64 cores of
a modern high-performance computing (HPC) cluster for the coarse
simulations and on the order of one week per cycle for the fine-grid
simulations.  We are currently working to develop effective multigrid
solvers for stable implicit time stepping schemes for the IB method
\cite{RDGuyXX-gmgiib}.  These methods promise to allow for
substantially larger time step sizes as well as shorter run times at
current grid resolutions.  They could also enable higher resolution
simulations that promise to resolve the fine-scale details of the
dynamics of the aortic root.

\section{Conclusions}

This work has presented new fluid-structure interaction models of the aortic root that
extend earlier IB models of the aortic valve by incorporating a
finite-strain model of the aortic root that uses a constitutive model
fit to tensile test data obtained from human aortic root tissue
specimens.  The models capture both the complex fluid dynamics of the
flow and the finite deformations of the vessel wall, and they yield
results that are in good agreement with the clinical data that were
used to fit the circulation model used in this study, as well as to
physiological data in the research literature.  In addition, a grid
convergence study demonstrates that nearly grid converged results are
obtained at practical spatial resolutions, although yet higher spatial
resolution is needed to obtain fully resolved simulation results.
Finally, by comparing the dynamics of symmetric and asymmetric aortic
root models, we find that the asymmetric model appears to yield a more
accurate description of both the bulk flow properties and also the
aortic wall mechanics.

\begin{acknowledgements}
  This work has been funded in part by American Heart Association
  award 10SDG4320049, National Institutes of Health awards GM071558
  and HL117063, and National Science Foundation awards DMS 1016554 and
  OCI 1047734.  VF was supported in part by the Polytechnic School of
  Engineering of New York University.  Computations were performed at
  New York University using computer facilities funded in large part
  by a generous donation by St.~Jude Medical, Inc.
\end{acknowledgements}

\bibliographystyle{spmpsci}       
\bibliography{bibliography}

\end{document}